\documentclass[floatfix,twocolumn,showpacs,aps,tightenlines]{revtex4-1}
\usepackage{graphicx}
\usepackage{color}
\usepackage{psfrag}
\usepackage{dcolumn}
\usepackage{bm}

\usepackage{amssymb}
\usepackage{amsmath}
\usepackage{amsfonts}
\usepackage{longtable}
\usepackage{xspace}
\usepackage{mathrsfs}

\newcommand{\beq}{\begin{equation}}
\newcommand{\eeq}{\end{equation}}

\newcommand{\be}{\begin{equation}}
\newcommand{\ee}{\end{equation}}
\newcommand{\bea}{\begin{eqnarray}}
\newcommand{\eea}{\end{eqnarray}}
\newcommand{\bes}{\begin{subequations}}
\newcommand{\ees}{\end{subequations}}

\newcommand{\scri}{\mathscr{I}}

\usepackage{bm}

\begin{document}

\title{The second RIT binary black hole simulations catalog
and its application to gravitational waves parameter estimation}

\author{James Healy}
\author{Carlos O. Lousto}
\author{Jacob Lange}
\author{Richard O'Shaughnessy}
\author{Yosef Zlochower}
\author{Manuela Campanelli}
\affiliation{Center for Computational Relativity and Gravitation,
School of Mathematical Sciences,
Rochester Institute of Technology, 85 Lomb Memorial Drive, Rochester,
New York 14623}

\date{\today}

\begin{abstract}
The RIT numerical relativity group is releasing the second public
catalog of black-hole-binary waveforms \url{http://ccrg.rit.edu/~RITCatalog}.
This release consists of 320 accurate simulations that include 46 precessing
and 274 nonprecessing binary systems with mass ratios $q=m_1/m_2$ in the
range $1/7\leq q\leq1$ and individual spins up to $s/m^2=0.95$.
The new catalog contains search and ordering tools
for the waveforms based on initial parameters of the binary, 
trajectory information, peak radiation, and final remnant black hole
properties. 
The final black hole remnant properties provided here can be used to model the
merger of black-hole binaries from its initial configurations.
The waveforms are extrapolated to 
future null infinity and can be used
to independently interpret
gravitational wave signals from laser interferometric detectors.
As an application of this waveform catalog we reanalyze the signal of GW150914
implementing parameter estimation techniques that make 
use of only numerical waveforms without any reference to information from
phenomenological waveforms models.
\end{abstract}

 \maketitle

\section{Introduction}\label{sec:Intro}

Ten years of advances and studies since the 
breakthroughs~\cite{Pretorius:2005gq,Campanelli:2005dd,Baker:2005vv} in
numerical relativity 
led to detailed predictions of the gravitational waves from the late inspiral,
plunge, merger, and ringdown of black-hole-binary systems (BHB).
These predictions helped to accurately identify the first direct
detection \cite{TheLIGOScientific:2016wfe} of gravitational waves with
such binary black hole systems \cite{Abbott:2016blz,Abbott:2016nmj,TheLIGOScientific:2016pea,Abbott:2016wiq} and match them to 
targeted supercomputer simulations \cite{Abbott:2016apu,TheLIGOScientific:2016uux,Lovelace:2016uwp}.
The observed gravitational waves were remarkably consistent with the predictions of
numerical general relativity \cite{TheLIGOScientific:2016src,TheLIGOScientific:2016pea,Abbott:2016wiq},
thereby supporting the notion that general
relativity is an accurate theory of gravity in the
highly dynamical, strong field regime of merging binary black holes.

Numerical relativity techniques have been used to explore the late
dynamics of spinning black-hole binaries, beyond the post-Newtonian
regime for several years. The first generic, long-term precessing
black-hole binary evolutions (i.e., without any symmetry) were
performed in Ref.~\cite{Campanelli:2008nk}, where a detailed
comparison with post-Newtonian $\ell=2,3$ waveforms was made. More
recently, the longest of such comparisons for a precessing binary
was done in \cite{Lousto:2015uwa} and a full
numerical simulation was performed for a nonspinning binary with 350 orbits
in Ref.~\cite{Szilagyi:2015rwa}.

Numerical simulations have started to explore the corners of parameter space,
these include near extremal \footnote{$\chi$ denotes the spin angular
momentum of a black hole in units of the square of its mass.  The
maximum possible spin is $\chi=1$} with $\chi=0.99$ spinning black-hole
binaries in Refs.~\cite{Lovelace:2014twa,Zlochower:2017bbg},
mass ratios as small as 
$q=1/100$ in Refs.~\cite{Lousto:2010ut,Sperhake:2011ik}, and large
initial separations, $R=100M$, in Ref.~\cite{Lousto:2013oza}.
Similarly challenging, high energy collision of black holes were
studied in Ref.~\cite{Sperhake:2008ga,Shibata:2008rq,Healy:2015mla} and hyperbolic black-hole
encounters in Ref.~\cite{Healy:2008js,Gold:2012tk}.

Other important studies include the exploration of the {\it hangup}
effect, i.e. the role individual black-hole spins play to delay or
accelerate their merger \cite{Campanelli:2006uy, Hannam:2007wf,Hemberger:2013hsa,Healy:2018swt}, the determination of the magnitude
and direction of the {\it recoil} velocity of the final merged black
hole \cite{Campanelli:2007ew,Campanelli:2007cga,Herrmann:2007ex,Pollney:2007ss,Baker:2006vn,Gonzalez:2007hi, Schnittman:2007ij,Lousto:2011kp},
and the {\it flip-flop} of individual spins during the orbital phase
\cite{Lousto:2014ida, Lousto:2015uwa, Lousto:2016nlp}, as well as
precession dynamics \cite{Schmidt:2010it,Lousto:2013vpa,Pekowsky:2013ska,Ossokine:2015vda,Lousto:2018dgd} and the inclusion
of those dynamics to construct surrogate models for gravitational
waveforms \cite{Blackman:2017dfb,Blackman:2017pcm,Varma:2018mmi}.

There have been several significant efforts to coordinate numerical
relativity simulations to support gravitational wave observations.
These include the numerical injection analysis (NINJA) project
\cite{Aylott:2009ya, Aylott:2009tn, Ajith:2012az, Aasi:2014tra}, the
numerical relativity and analytical relativity (NRAR) collaboration
\cite{Hinder:2013oqa}, and the waveform catalogs released by the
SXS collaboration~\cite{Mroue:2013xna,Blackman:2015pia, Chu:2015kft},
Georgia Tech.~\cite{Jani:2016wkt}, and RIT~\cite{Healy:2017psd}. 

In this paper we describe a new release of the public waveform catalog
by the RIT numerical relativity group that nearly triples the
number of waveforms by adding a new set of 194 waveforms,
with 154 aligned spins and 40 precessing binaries.
The catalog has new search and ordering features
and includes all modes $\ell\leq4 $ modes of $\psi_4$ and the strain $h$
(both extrapolated to null-infinity).  The catalog can be accessed
from \url{http://ccrg.rit.edu/~RITCatalog}.

This paper is organized as follows.  In Section \ref{sec:FN} we
describe the methods and criteria for producing the numerical
simulations and evaluation of their errors in order to be included in
the RIT catalog.  In Sec.~\ref{sec:Catalog}
we describe the use of the new searching and ordering capabilities of
all the relevant BHB parameters, the file format, and the  full content
of the data in the catalog.  In Sec.~\ref{sec:GW150914} we use the
waveform catalog to estimate the binary black hole parameters that
best match the first gravitational wave event GW150914.  We use the
Bayesian likelihood maximized over extrinsic
parameters as described in \cite{Lange:2017wki}
and map it onto the grid of simulations. Use of interpolation routines
lead to an estimate of the confidence intervals that are consistent
with previous estimates for the aligned spin binaries.  We conclude in
Sec.~\ref{sec:Discussion} with a discussion of the future use of this
catalog for parameter inference of new gravitational waves events and
the extensions to this work to more generic precessing binaries.

\section{Full Numerical Evolutions}\label{sec:FN}

The simulations in the RIT Catalog were evolved using the {\sc
LazEv} code~\cite{Zlochower:2005bj} implementation of the moving puncture
approach~\cite{Campanelli:2005dd} (with the modifications suggested by
Ref.~\cite{Marronetti:2007wz}). In all cases (except the very high spin
where we use CCZ4~\cite{Alic:2011gg}) we use the BSSNOK
(Baumgarte-Shapiro-Shibata-Nakamura-Oohara-Kojima) family
of evolutions systems~\cite{Nakamura87, Shibata95, Baumgarte99}.
For the runs in the catalog, we used a variety of finite-difference
orders, Kreiss-Oliger dissipation orders, and Courant factors~\cite{Lousto:2007rj,Zlochower:2012fk,Healy:2016lce}.
All of these are
given in the metadata included in the catalog and the references associated
with each run (where detailed studies have been performed). 

The {\sc LazEv} code uses the {\sc EinsteinToolkit}~\cite{Loffler:2011ay, einsteintoolkit} / {\sc Cactus}~\cite{cactus_web} / {\sc Carpet}~\cite{Schnetter-etal-03b}
infrastructure.  The {\sc Carpet} mesh refinement driver provides a
``moving boxes'' style of mesh refinement. In this approach, refined
grids of fixed size are arranged about the coordinate centers of both
holes.  The code then moves these fine grids about the computational
domain by following the trajectories of the two black holes (BHs).

We use {\sc AHFinderDirect}~\cite{Thornburg2003:AH-finding} to locate
apparent horizons.  We measure the magnitude of the horizon spin using
the {\it isolated horizon} (IH) algorithm detailed in
Ref.~\cite{Dreyer02a} (as  implemented in
Ref.~\cite{Campanelli:2006fy}).
Once we have the horizon spin, we can calculate the horizon
mass via the Christodoulou formula 
${m_H} = \sqrt{m_{\rm irr}^2 + S_H^2/(4 m_{\rm irr}^2)}\,,$
where $m_{\rm irr} = \sqrt{A/(16 \pi)}$ and  $A$ is the surface area
of the horizon. 

To compute the numerical initial data, we use the puncture
approach~\cite{Brandt97b} along with the {\sc
  TwoPunctures}~\cite{Ansorg:2004ds} code.  To compute initial low
eccentricity orbital parameters, we use the post-Newtonian techniques
described in~\cite{Healy:2017zqj} to determine quasi-circular orbits.
We then evaluate the residual
eccentricity during evolution via the simple formula,
as a function of the separation of the holes, $d$,
$e_d=d^2\ddot{d}/M$, as given in \cite{Campanelli:2008nk}.


As discussed in Ref. \cite{Healy:2017psd} the main sources
of numerical errors in this catalog are 
due to finite difference truncation, finite extraction
radii, finite number of modes, and 
the non-zero residual initial eccentricities and displacement
of the center of mass.

During the early inspiral, the irreducible masses and intrinsic spins
of each black hole should be nearly constant because the levels of
gravitational
wave energy and momentum absorbed by the holes is 4-5 orders of
magnitude smaller\cite{Isoyama:2017tbp} than those emitted to
infinity. During a simulation, the masses and spins vary due to
numerical truncation error, and we then use these variations as a
measure of the size of the truncation error.
For our current simulations we monitor accuracy by measuring
the conservation of the individual horizon masses and spins during evolution,
as well as the level of satisfaction of the Hamiltonian and momentum constraints,
to ensure reaching an accuracy consistent with our main applications.
Those measurements are seen to be preserved at least to one 
part in $10^{4}$ in the cases of the masses and one part in $10^3$ in the cases of the spins (see for instance Fig. 6 in Ref.~\cite{Lousto:2018dgd}). 

We measure radiated energy,
linear momentum, and angular momentum, in terms of the radiative Weyl
Scalar $\psi_4$, using the formulas provided in
Refs.~\cite{Campanelli:1998jv, Lousto:2007mh}. 
These formulas are
strictly speaking only valid on future null-infinity ($\scri^+$).
We therefore measure the radiated energy-momentum on a series of
timelike worldtubes  of finite
 radius and then extrapolate to $r=\infty$ using both linear and
quadratic extrapolations. The difference between these two
extrapolations is an estimate for the uncertainty.

Unlike the radiated energy-momentum, more care is needed to properly
extrapolate the waveform itself to $\scri^+$.
As described in
Ref.~\cite{Nakano:2015pta},
we use the Teukolsky equation to obtain expressions for $r \psi_4$ at
$\scri^+$
based on its values on a timelike worldtube traced out by a fixed sphere of
constant (large) areal radius $r$ [see Eq. (29), there].
The expressions there contain the corrections of order ${\cal O}(1/r)$
and ${\cal O}(1/r^2)$ to $r \psi_4$. As shown in
Ref.~\cite{Nakano:2015pta}, this extrapolation is consistent with both
the waveform and the radiated energy-momentum extrapolated
using a least squares fit to a polynomial in $1/r$.
Additionally, the ${\cal O}(1/r)$
perturbative corrections were shown to be consistent with a
Cauchy-Characteristic extraction for an equal-mass binary
in~\cite{Babiuc:2010ze}.

Various simulations in this catalog were studied in
detail in previous papers. 
In Appendix A of Ref.~\cite{Healy:2014yta}, we
performed a detailed error analysis of configurations with
equal mass and spins aligned/antialigned with respect to the orbital
angular momentum; 
in Appendix B of Ref.~\cite{Healy:2016lce}, we performed
convergence studies for runs with mass ratios $(q=1,3/4,1/2,1/3)$ and
measured errors due to finite observer locations; 
and in Ref.~\cite{Healy:2017mvh}, we performed convergence studies
for $q\geq 1/10$ nonspinning binaries.

Finally, in addition to all the internal consistency analysis and
error estimates, in Ref.~\cite{Lovelace:2016uwp} we showed that for
the parameter estimated for GW150914 ($q=m_1/m_2=0.82$ and spins for
the small/large holes of $\chi_1=-0.44$ and $\chi_2=+0.33$), the RIT
waveforms and those produced completely independently by the SXS
collaboration have an excellent match~\cite{Cho:2012ed} of $\gtrsim 0.99$ overall
for modes up to $\ell=5$. In
Ref.~\cite{Healy:2017abq} a similar agreement between approaches has
been found for five targeted precessing and nonprecessing simulations
of GW170104, displaying a 4th order convergence with finite difference
resolution. The comparisons were also carried up to $\ell=5$-modes.
For all modes up to $l\leq4$ we found a match of $\gtrsim 0.99$ 
and $\gtrsim 0.97$ for the $l=5$ modes.

In all our studies we concluded that the waveforms at the resolutions
provided in this catalog are well into the convergence regime (roughly
converging at 4th-order with resolution), that the horizon evaluated
quantities such as the remnant final mass and spins have errors of the
order of 0.1\%, and that the radiatively computed quantities such as
the recoil velocities and peak luminosities are evaluated at a typical
error of 5\%.

\section{The Catalog}\label{sec:Catalog}

\begin{figure}
  \includegraphics[angle=270,width=0.494\columnwidth]{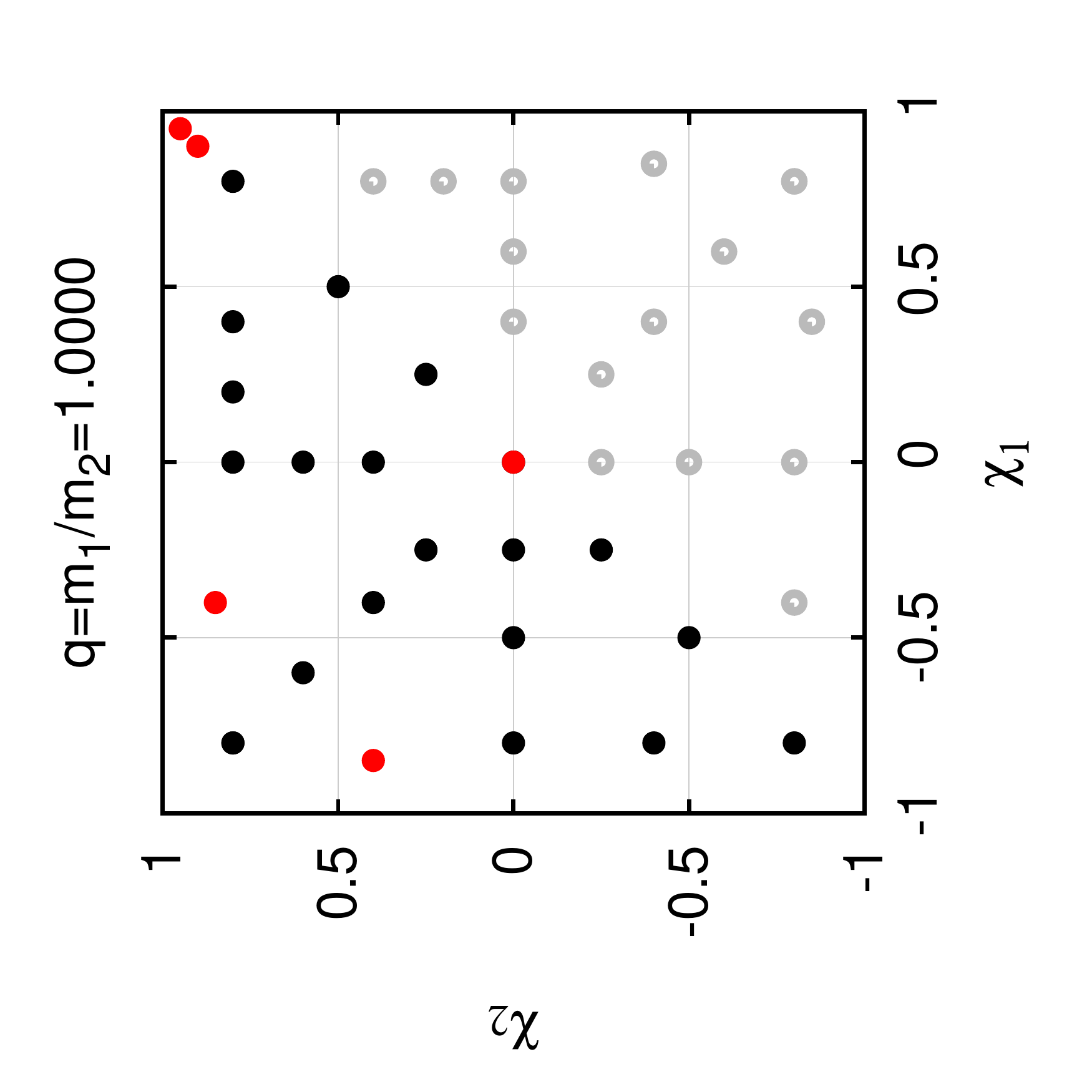}
  \includegraphics[angle=270,width=0.494\columnwidth]{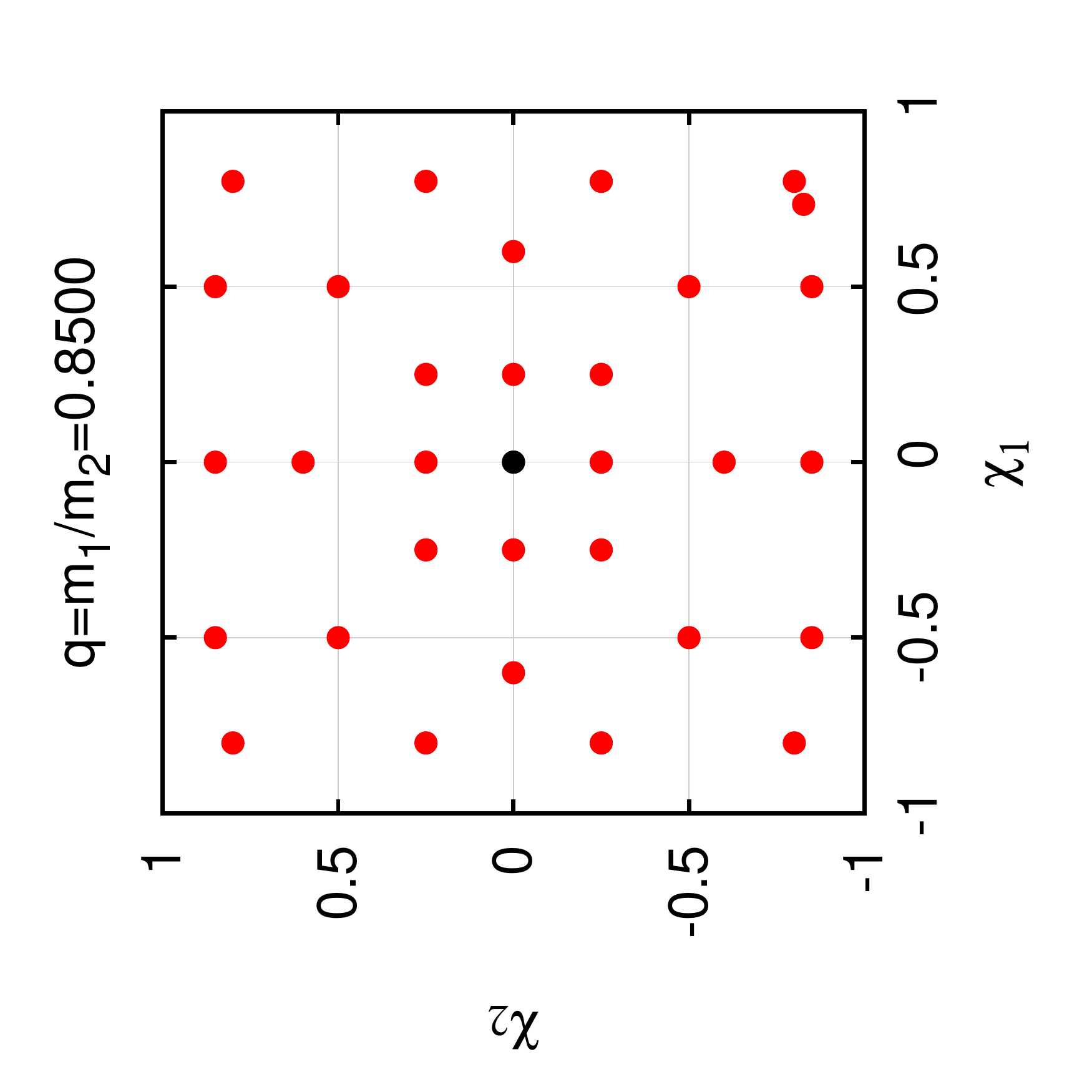}\\
    \includegraphics[angle=270,width=0.494\columnwidth]{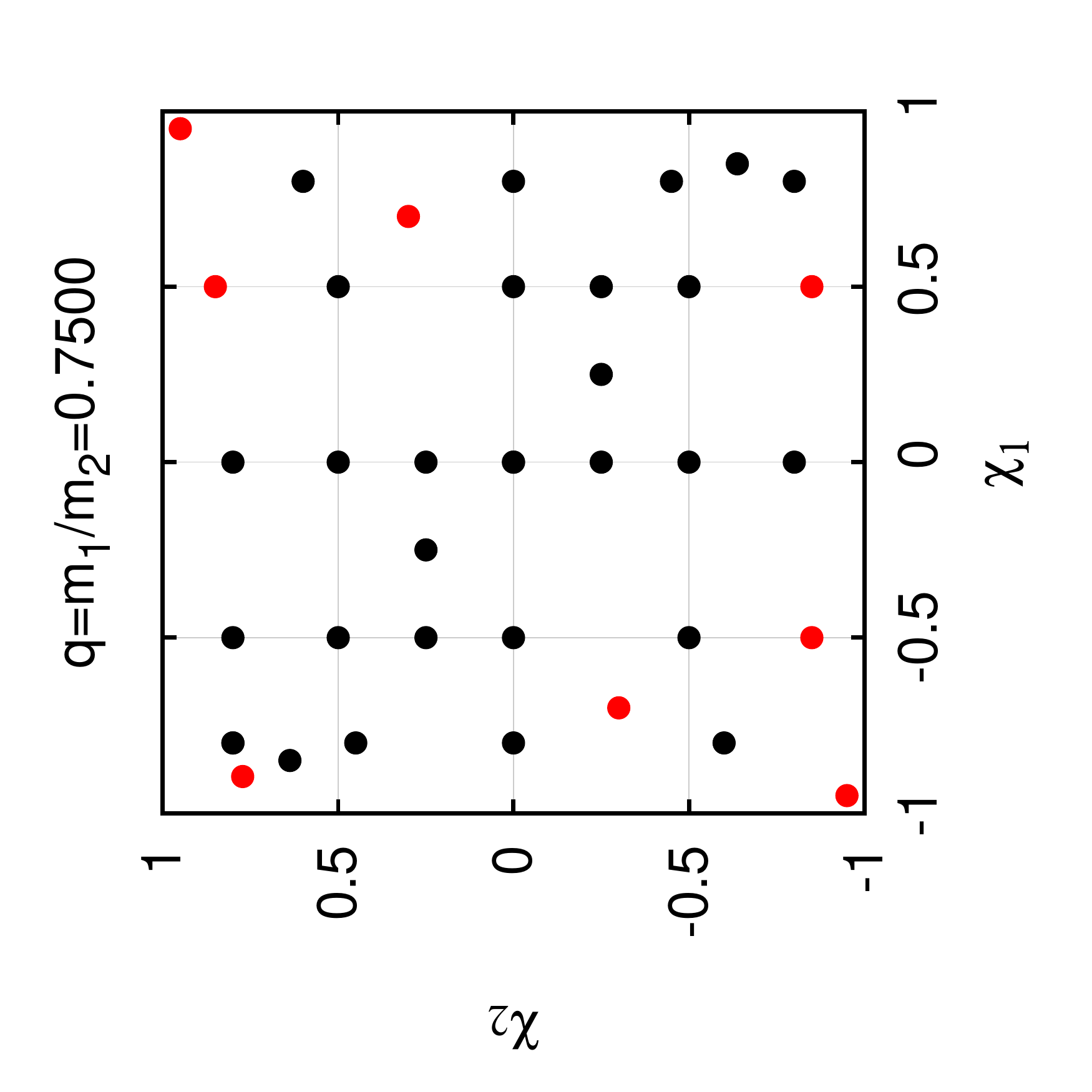}
  \includegraphics[angle=270,width=0.494\columnwidth]{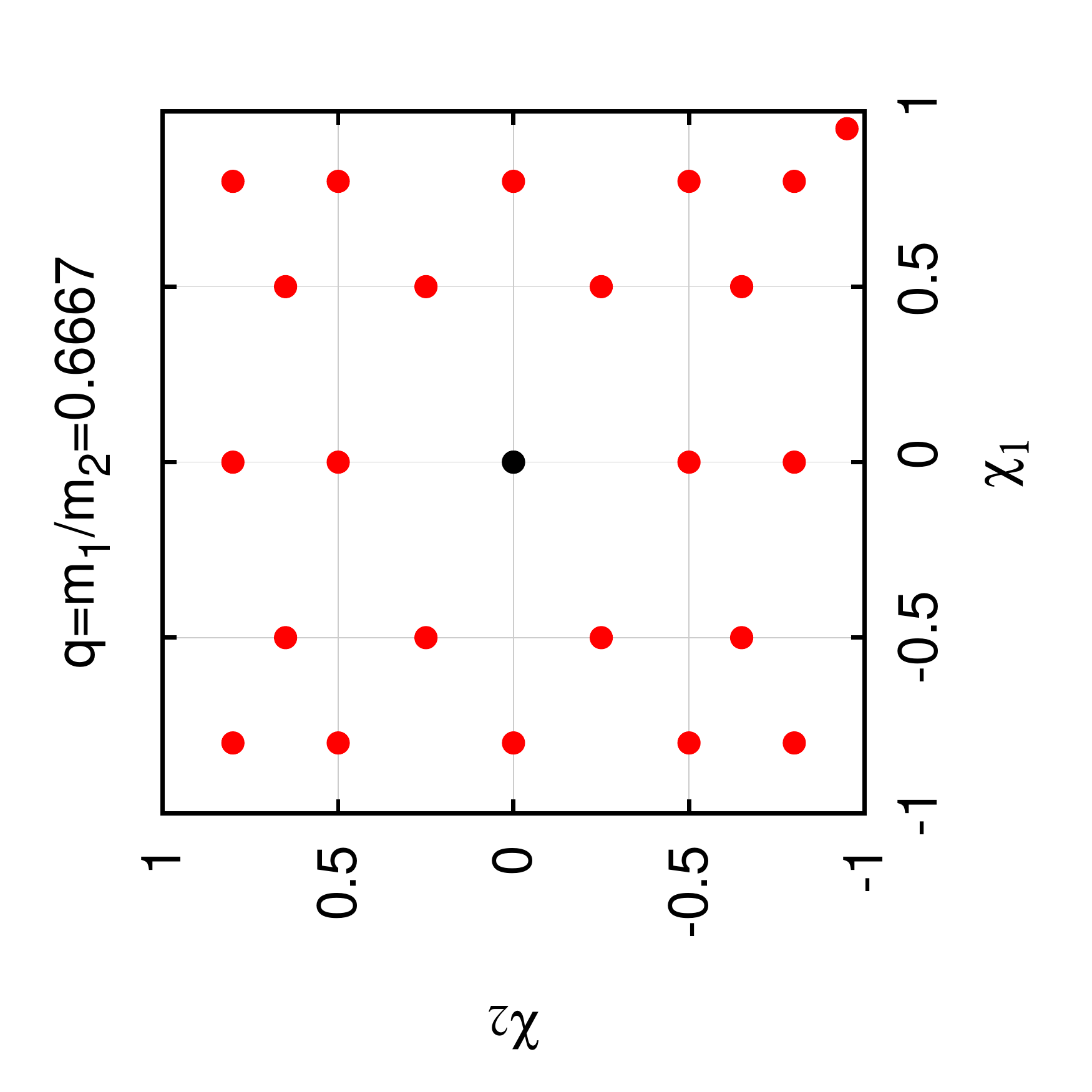}\\
    \includegraphics[angle=270,width=0.494\columnwidth]{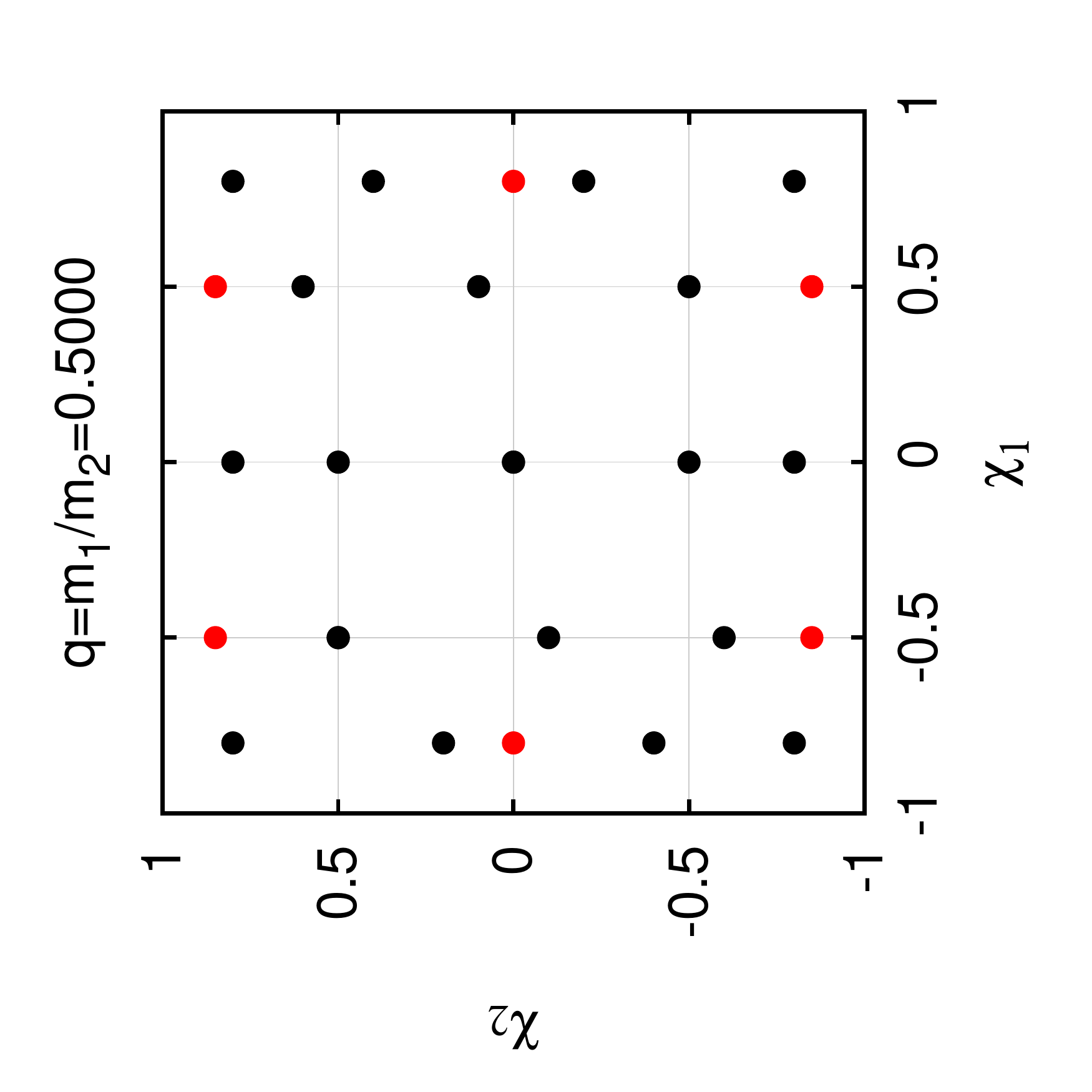}
  \includegraphics[angle=270,width=0.494\columnwidth]{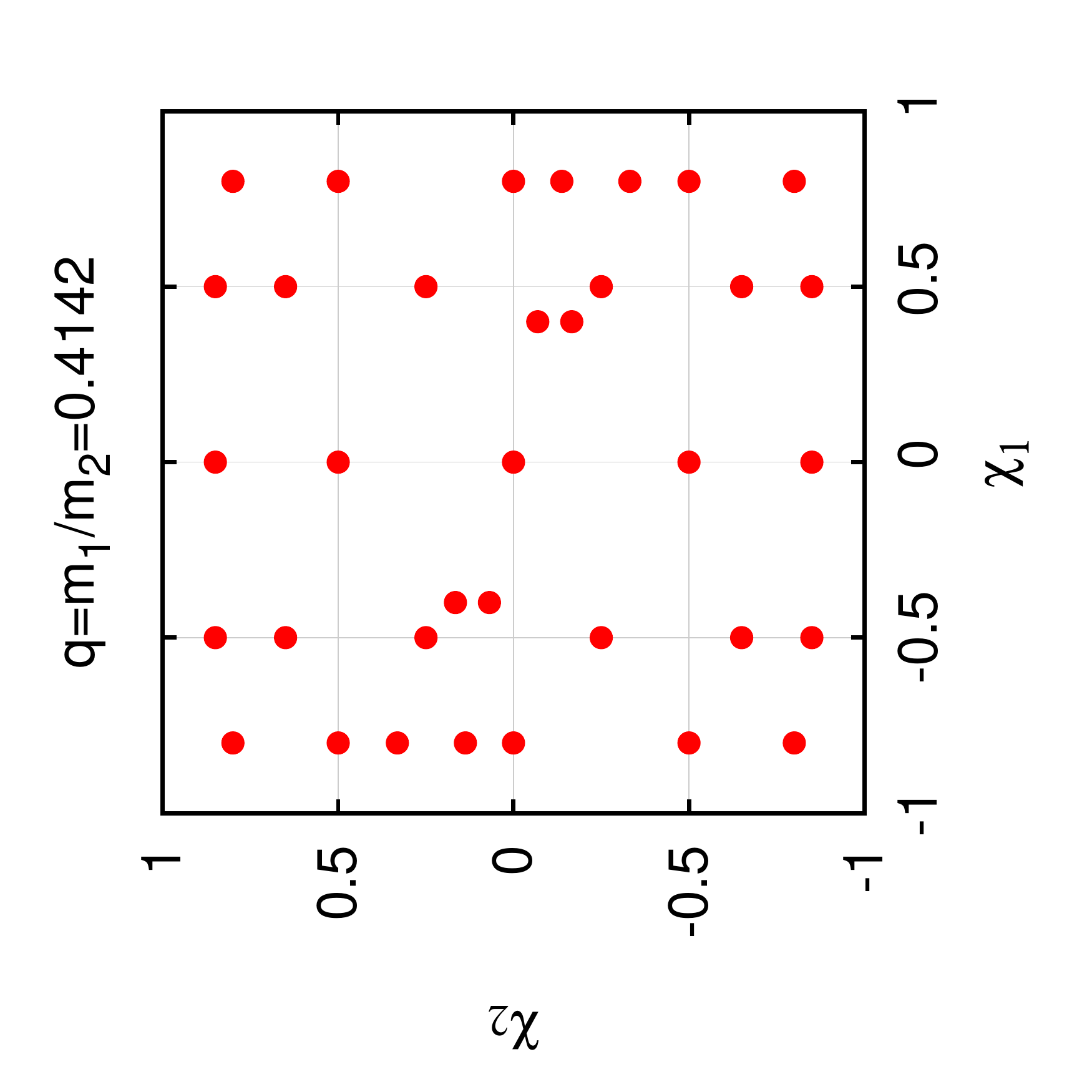}\\
    \includegraphics[angle=270,width=0.494\columnwidth]{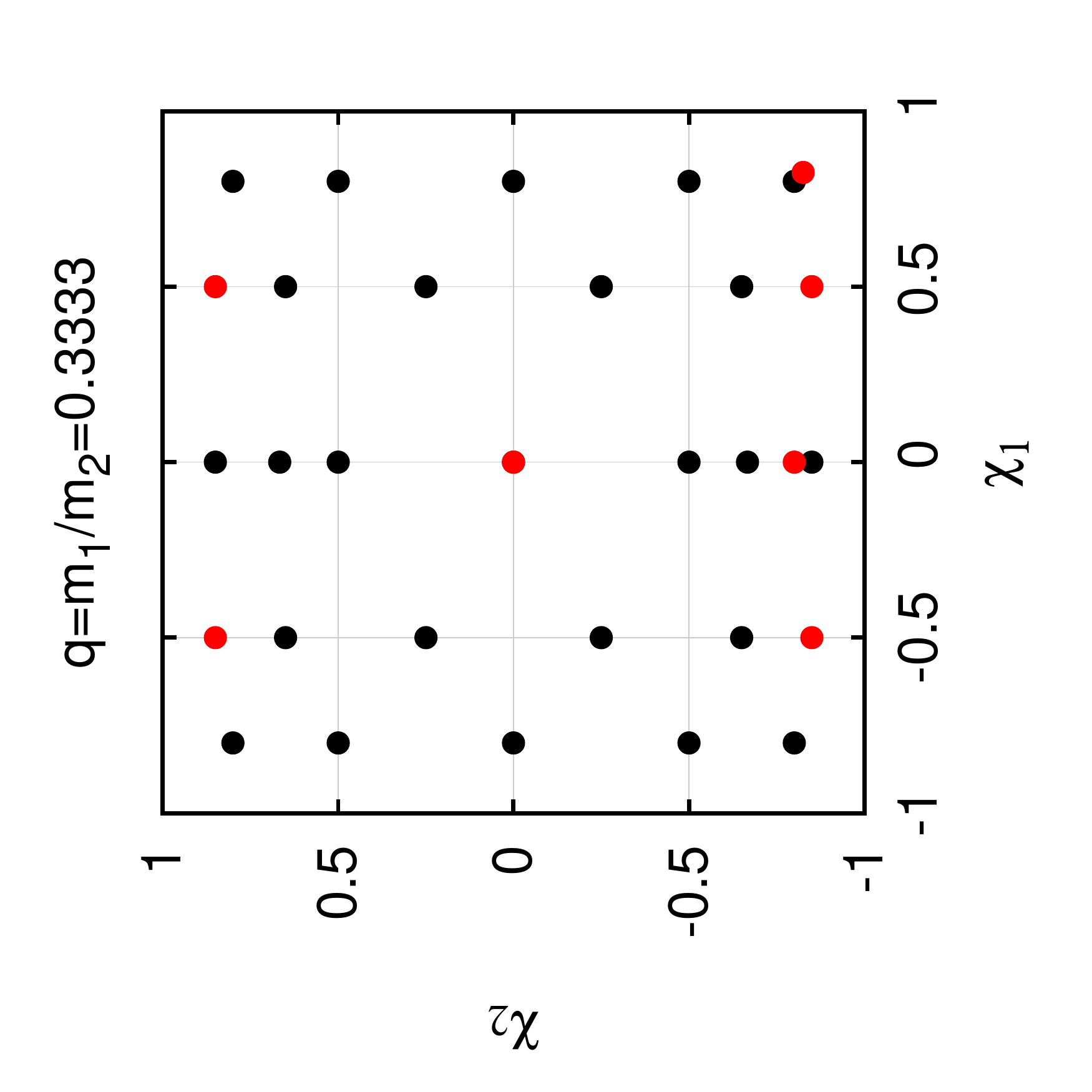}
  \includegraphics[angle=270,width=0.494\columnwidth]{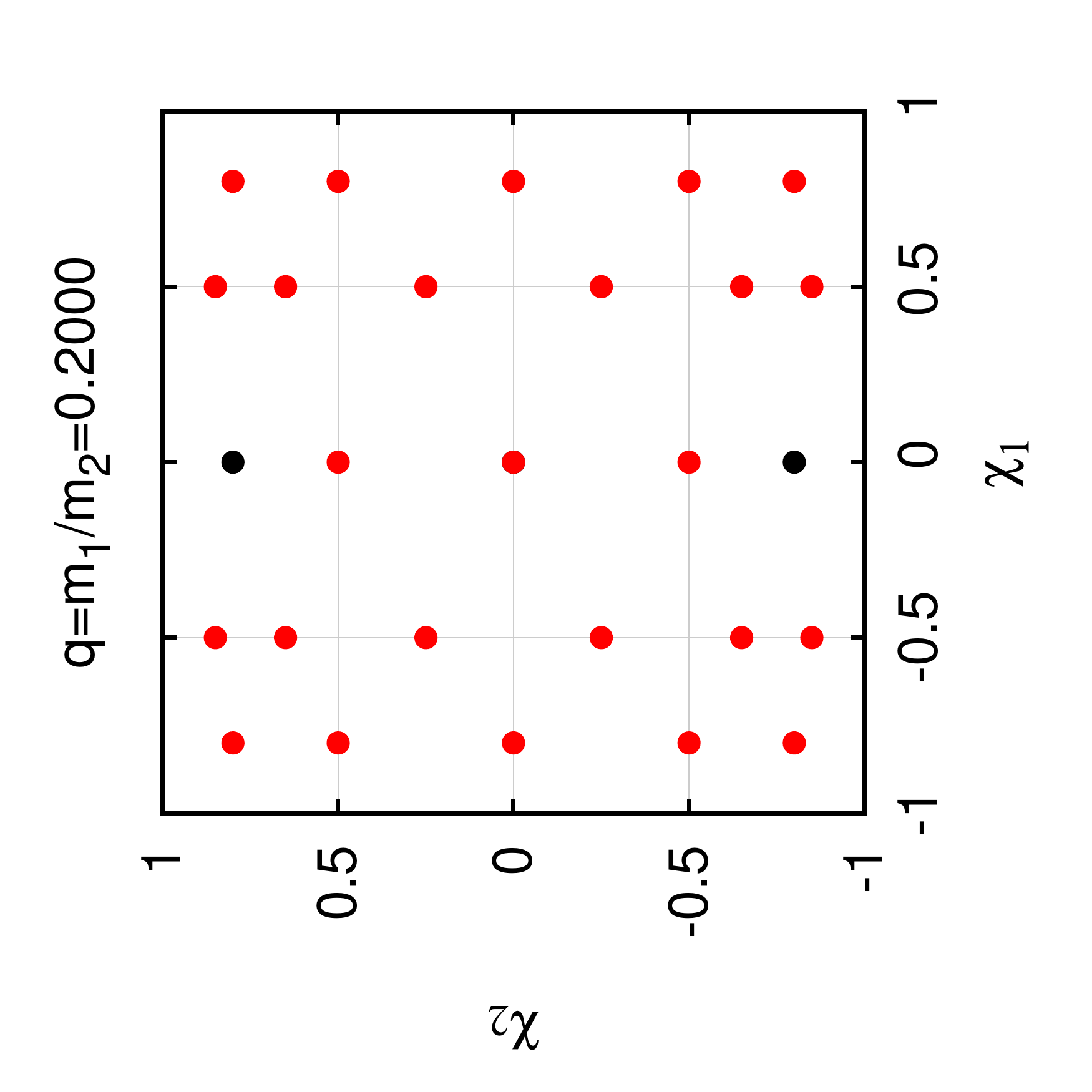}
  \caption{Initial parameters in the $(q,\chi_1,\chi_2)$ space
    for the 274 nonprecessing binaries. Note that $\chi_i$ denotes 
    the component of the dimensionless spin of BH $i$ along the
    orbital angular momentum.
    Each panel corresponds to a given mass ratio that covers the 
    comparable masses binary range from $q=1$ to $q=1/5$.
    The dots in black denote the simulations of the catalog first
    release, and the dots in red are those of this second release.
    \label{fig:panels}}
\end{figure}

\begin{figure}
  \includegraphics[angle=0,width=1.0\columnwidth]{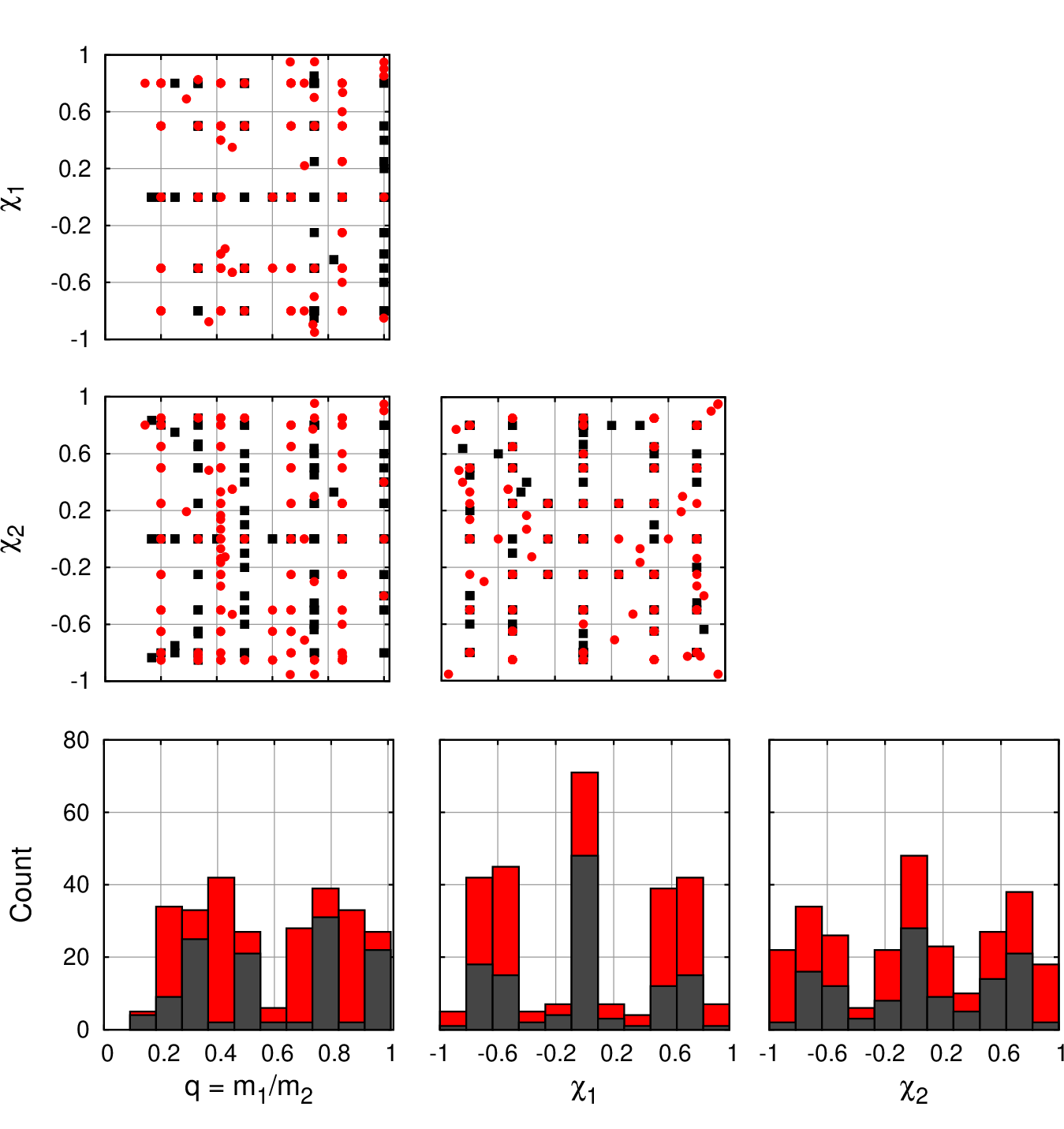}
  \caption{Counting simulations in the $(q,\chi_1,\chi_2)$ planes 
    (faces of the cube) for the 274 nonprecessing binaries.  The 120 release 1
    simulations are black and the 154 release 2 simulations are red.
      \label{fig:AlignedMulti}}
\end{figure}

\begin{figure}
  \includegraphics[width=0.7\columnwidth]{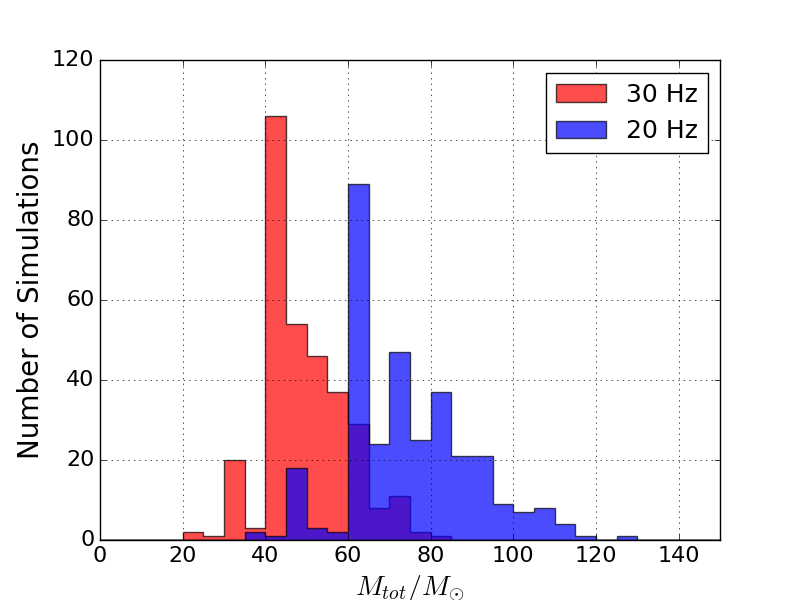}\\
\includegraphics[width=0.7\columnwidth]{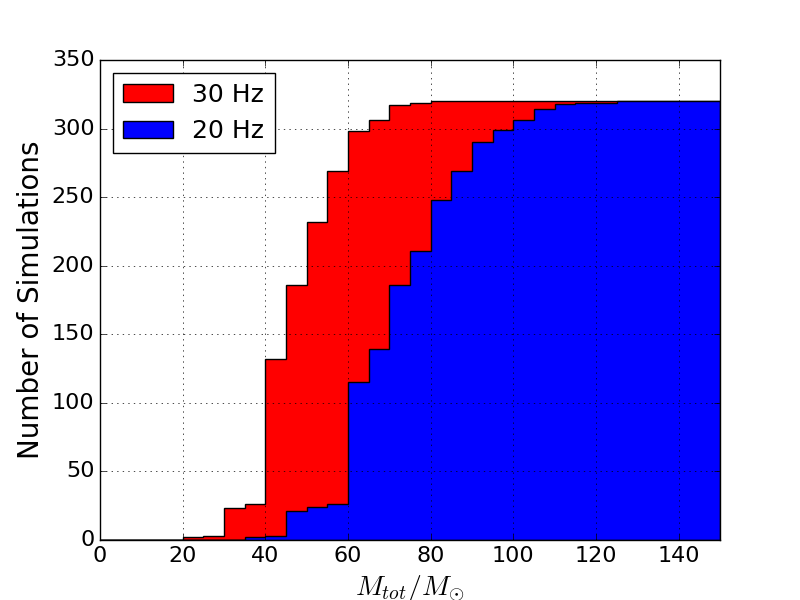}
  \caption{Top: Distributions of the total mass of BHB systems in the
RIT catalog corresponding to a starting gravitational wave frequency of
20 Hz (blue) and 30 Hz (red) in bins of $5M_\odot$. 
Bottom: The cumulative version of the above plot also in bins of
$5M_\odot$ for the 320 simulations in this catalog.
\label{fig:mtotal}}
\end{figure}

The RIT Catalog can be found at \url{http://ccrg.rit.edu/~RITCatalog}.
Figure~\ref{fig:panels} shows the distribution of the
non-precessing runs in the catalog in
terms of $\chi_{1,2}$ and $q$ (where $\chi_i$ is the component of
the dimensionless spins of BH $i$ along the direction of the orbital angular
momentum).
The information currently in the catalog consists of the metadata
describing the runs and all modes up through the $\ell=4$ modes
(enough for most applications) of $M r \psi_4$ 
extrapolated to $\scri^+$ via the perturbative approach
of~\cite{Nakano:2015pta}. 
The associated metadata include the
initial orbital frequencies, ADM masses, initial waveform frequencies from (2,2) mode,  black hole masses, momenta, spins, separations,
and eccentricities, as well the black-hole masses and spins once the
initial burst of radiation has left the region around the binary.
Note that we normalize our data such that
the sum of the two initial horizon masses is $1M$.
These {\it relaxed} quantities (at $t_{relax}=200M$ after the initial burst
of radiation has mostly dissipated)
are more accurate and physically relevant for modeling purposes. 
In addition, we also include the final remnant black hole masses,
spins and recoil velocity.

The catalog is organized using an interactive table~\cite{datatables_web} that includes an
identification number, resolution, type of run (nonspinning, aligned spins,
precessing), the initial proper length of the coordinate 
line joining the two BH centroids that is outside both horizons~\cite{Lousto:2013oza},
the coordinate separation of the two centroids, the mass ratio of the two black
holes, the components of the dimensionless spins of the two black
holes, the starting waveform frequency, $M f_{22, {\rm relax}}$, 
time to merger, number of gravitational wave cycles 
calculated from the (2,2) modes from the beginning of the inspiral signal 
to the amplitude peak,
remnant mass, remnant spin, recoil
velocity, peak luminosity, amplitude and frequency.
The final column gives the appropriate
bibtex keys for the relevant publications where the waveforms were
first presented. The table can be sorted (ascending or descending) by
any of these columns. And there is a direct search feature that runs over all
table elements.

Resolutions are given in terms of the grid spacing of the refinement
level where the waveform is extracted (which is typically two
refinement levels below the coarsest grid) with $R_{obs}\sim100M$.
We use the notation 
nXYY, where the grid spacing in the wavezone is given by $h=M/X.YY$, e.g.,
n120 corresponds to $h=M/1.2$.

For each simulation in the catalog there are three files: one contains
the metadata information in ASCII format, the other two are a tar.gz files
containing ASCII files with up to and including $\ell=4$
modes of $M r\psi_4$ and $h$.  In the near future, data will be available in the
Numerical Relativity Injection format \cite{Schmidt:2017btt}.
Note that the primary data in our catalog is the Weyl scalar $Mr \psi_4$ extrapolated
to $\scri^+$  (using Eq.~(29) of Ref.~\cite{Nakano:2015pta}),
rather than the strain $(r/M)h$.
We provide the strain but also leave it to the user to
convert $Mr\psi_4$ to strain for most modes
since this is still a sensitive process and is best handled on a
mode-by-mode basis.
The subtleties associated with transforming $\psi_4$ to $h$ arise from
the two integrations required \cite{Baker:2002qf,Campanelli:2008nk}.
One of the standard techniques, developed in Ref.~\cite{Reisswig:2010di},
performs this integration in Fourier space with a windowing function
and a low-frequency cutoff. Both of these require fine-tuning of
parameters. The codes to do this are open-source and publicly 
available from \url{https://svn.einsteintoolkit.org/pyGWAnalysis/trunk}.

Figure~\ref{fig:AlignedMulti} shows the distribution of the 274 non-precessing
runs in the catalog in terms of $\chi_{1,2}$ and $q$.
Those runs were  motivated by
systematic studies to produce a set of accurate remnant
formulas to represent the final mass, spin and recoil of a merged
binary black hole system and the peak Luminosity, amplitude and
frequency, as a function of the parameters of the
precursor binary, as reported in 
\cite{Healy:2014yta,Healy:2016lce,Healy:2018swt}.
A second important motivation was to provide a grid of simulations for
parameter estimation of gravitational wave signals detected by LIGO
using the methods described in \cite{Abbott:2016apu}. We will see
in the next section that we have achieved a good coverage of this
BHB parameter space.

The precessing runs in the catalog were motivated to study  
particular spin dynamics of merging BHB, such as
the study of unstable spin flip-flop,
as reported in \cite{Lousto:2016nlp} and the targeted followups
of gravitational wave signal from the first and second LIGO observing runs
\cite{Lovelace:2016uwp,Healy:2017abq}.
We have also payed special attention to the systematic study of simulations
covering a 4-dimensional parameter space involving a spinning and a nonspinning black hole binary as a function of the mass ratio.
Those simulations were originally performed to study remnant recoil and final masses and spins \cite{Zlochower:2015wga}. We have supplemented them here with
additional 31 simulations to have a coverage of spin orientations
(see Table \ref{tab:ID} that allows an
estimation of precession as shown in Fig.~\ref{fig:heatmapprec}).

Figure~\ref{fig:mtotal} shows the distributions of the minimal total mass of the 
BHB systems in the catalog given a starting gravitational wave frequency
of 20 or 30 Hz in the source frame.
This provides a coverage for the current events observed by LIGO (redshift
effects improve this coverage by a factor of $1+z$, where $z$ is the
redshift). Coverage
of even lower total masses would require longer simulations or hybridization
of the current waveforms with Post-Newtonian methods~\cite{Ajith:2012az}.

\section{Application of the Catalog to parameter estimation of Binary Black holes}\label{sec:GW150914}

We can directly compare any of our simulations to real or synthetic gravitational wave observations by scaling that
simulation and its predictions to a specific total redshifted mass $M_z$ and then marginalizing the likelihood for the gravitational
wave data over all extrinsic parameters
\cite{2015PhRvD..92b3002P,Abbott:2016apu,2017PhRvD..96j4041L,2017CQGra..34n4002O,2018arXiv180510457L}: 
 the seven coordinates characterizing the spacetime coordinates and orientation of the binary relative to the earth.  
Specifically the likelihood of the data given Gaussian noise has the form  (up to normalization)
\begin{equation}
\label{eq:lnL}
\ln {\cal L}(\bm{\lambda} ;\theta )=-\frac{1}{2}\sum\limits_{k}\langle h_{k}(\bm{\lambda} ,\theta )-d_{k} |h_{k}(\bm{\lambda} ,\theta )-d_{k}\rangle _{k}-\langle d_{k}|d_{k}\rangle _{k},
\end{equation}
where $h_{k}$ are the predicted response of the k$^{th}$ detector due to a source with parameters ($\bm{\lambda}$, $\theta$) and
$d_{k}$ are the detector data in each instrument k; $\bm{\lambda}$ denotes the combination of redshifted mass $M_{z}$ and the
remaining intrinsic parameters (mass ratio and spins; with eccentricity~$\approx0$)
needed to uniquely specify the binary's dynamics; $\theta$ represents the
seven extrinsic parameters (4 spacetime coordinates for the coalescence event and 3 Euler angles for the binary's
orientation relative to the Earth); and $\langle a|b\rangle_{k}\equiv
\int_{-\infty}^{\infty}2df\tilde{a}(f)^{*}\tilde{b}(f)/S_{h,k}(|f|)$ is an inner product implied by the k$^{th}$ detector's
noise power spectrum $S_{h,k}(f)$. 
In practice we adopt a low-frequency cutoff f$_{\rm min}$ so all inner products are modified to
\begin{equation}
\label{eq:overlap}
\langle a|b\rangle_{k}\equiv 2 \int_{|f|>f_{\rm min}}df\frac{[\tilde{a}(f)]^{*}\tilde{b}(f)}{S_{h,k}(|f|)}.
\end{equation}
For our analysis of GW150914, we adopt the same noise power spectrum employed in previous work 
\cite{Abbott:2016apu,2018arXiv180510457L}.
After exploring a range of redshifted masses $M_z$ for each simulation, we estimate $\ln {\cal L}(M_z)$ as a
function of mass for that
simulation and thus in particular its maximum value $\ln {\cal L}_{\rm max}$, as in \cite{Abbott:2016apu,2017PhRvD..96j4041L}.




\subsection{Non-precessing binaries}

\begin{widetext}
\begin{figure*}
\includegraphics[angle=0, width=0.75\columnwidth]{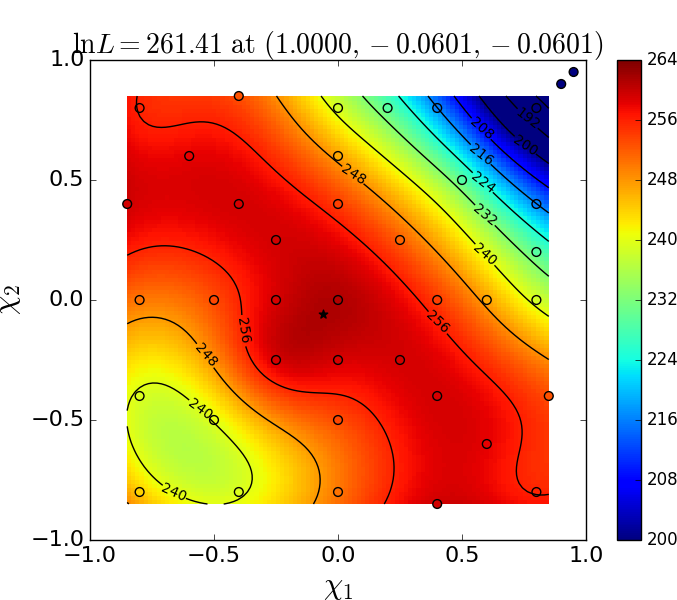}
\includegraphics[angle=0, width=0.75\columnwidth]{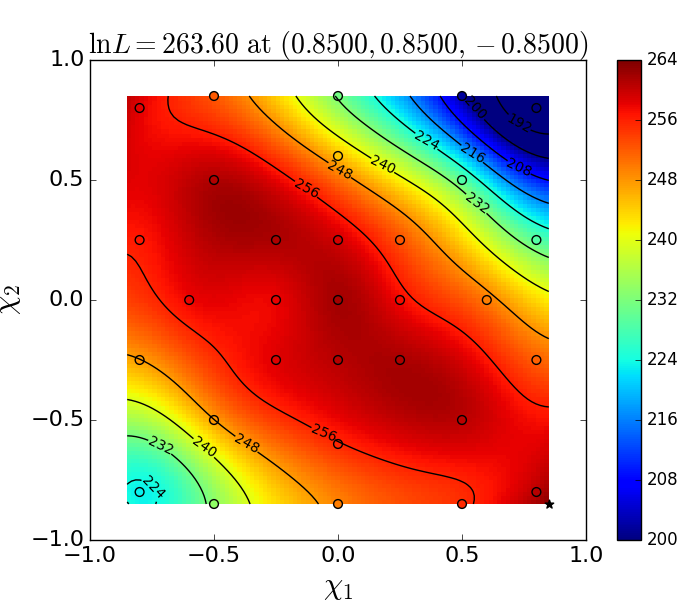}\\
\includegraphics[angle=0, width=0.75\columnwidth]{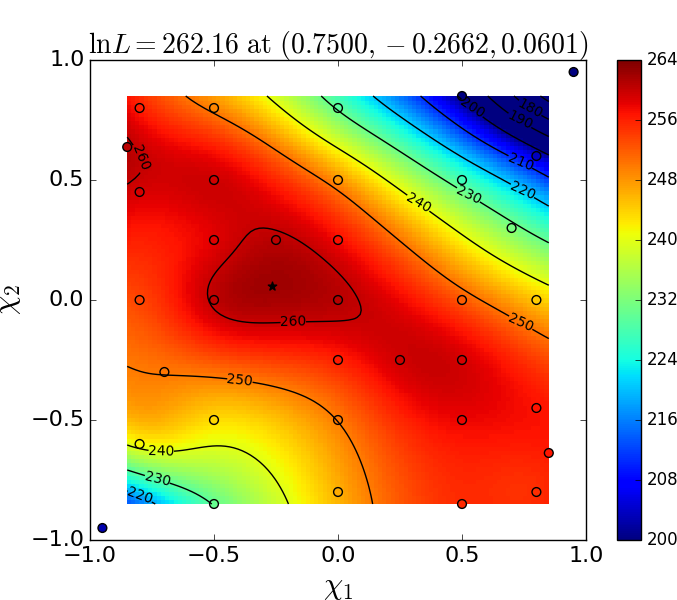}
\includegraphics[angle=0, width=0.75\columnwidth]{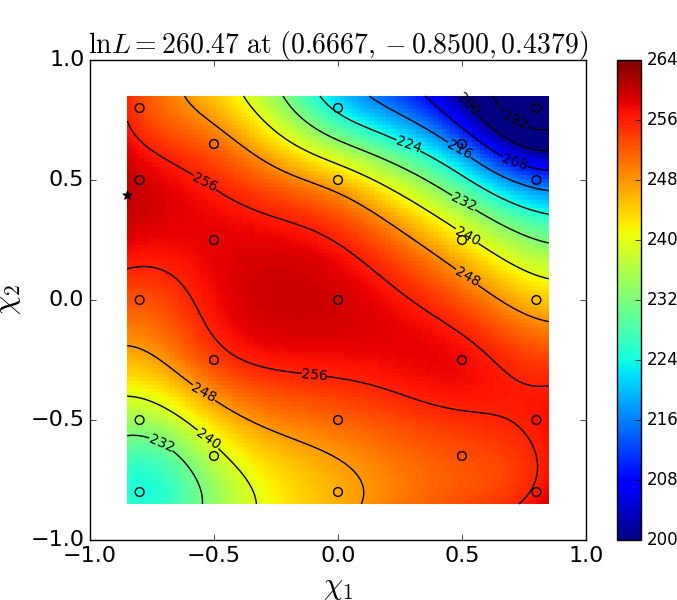}\\
\includegraphics[angle=0, width=0.75\columnwidth]{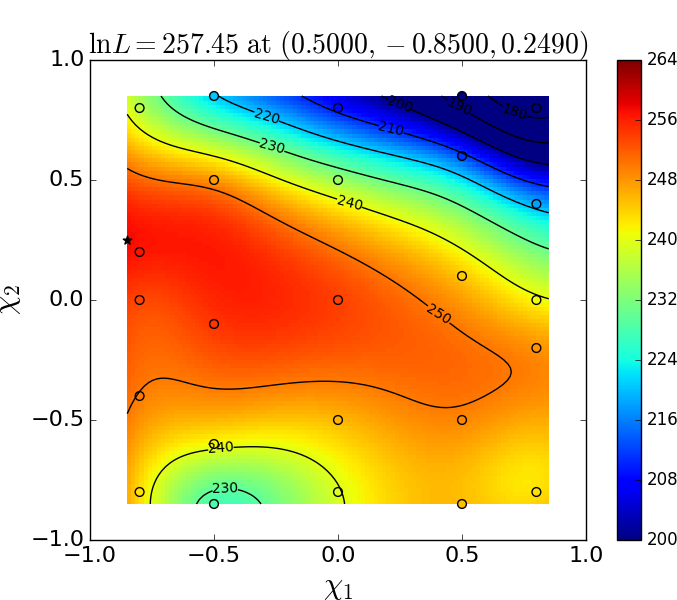}
\includegraphics[angle=0, width=0.75\columnwidth]{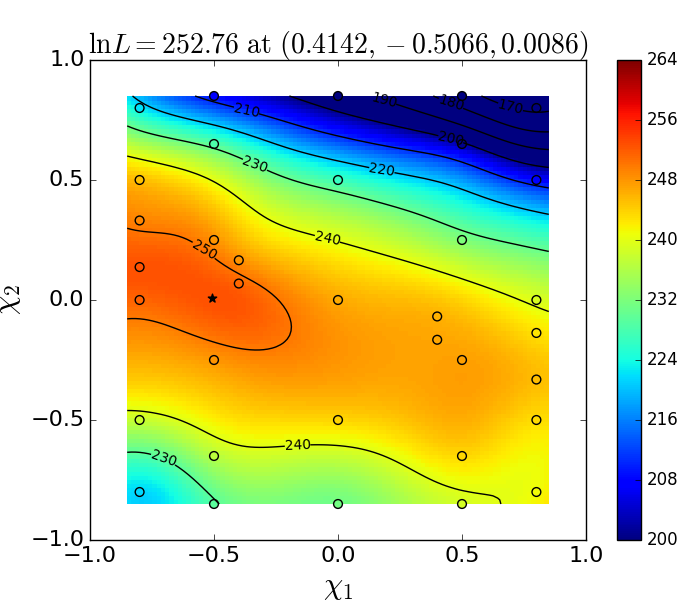}\\
\includegraphics[angle=0, width=0.75\columnwidth]{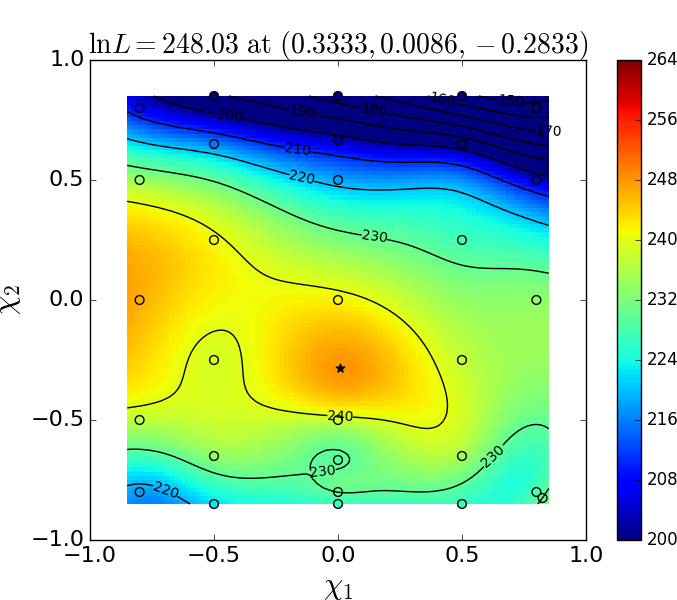}
\includegraphics[angle=0, width=0.75\columnwidth]{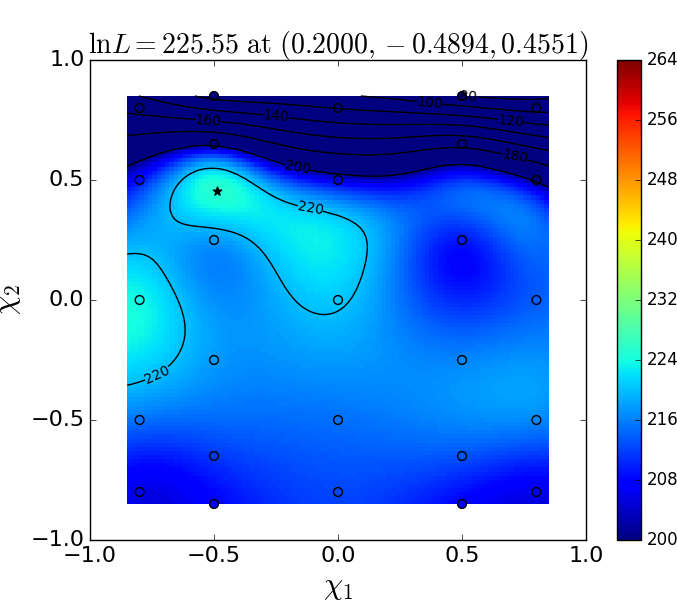}
  \caption{Heat maps of the GW150914 likelihood for each of the eight mass ratio panels covering form $q=1$ to $q=1/5$ and aligned/antialigned individual spins.
The individual panel with $q=0.85$ contains the highest likelihood.  Contour lines are in increments of 5.  The interpolated
    $\ln {\cal L}$ maximum at its location in $(q,\chi_1,\chi_2)$ space is given in each panel's title and denoted by the *
    in the plots.
\label{fig:heatmap}}
\end{figure*}
\end{widetext}

Fig.~\ref{fig:heatmap} displays a likelihood map for the simulations as a function of black hole's individual spins, with a panel for each of
the eight mass ratios studied, $q=1.00,0.85,0.75,0.6667,0.4142,0.50,0.3333,0.20$ on both LIGO detectors, H1 and L1, combined.
Our priors of intrinsic parameters are the discrete set of numerical relativity simulations ($q,\chi_1,\chi_2$),
with 100 points in uniform spacing between $40 < M_{total}/M_{\odot} < 120$.
The heat maps are generated using a multiquadric radial basis
interpolating function $\sqrt{(D/\epsilon)^2 + 1}$ between
the computed likelihoods for each simulation plotted by hollow circles
($D$ being the distance of the point in parameter space and $\epsilon=0.25$).
We have options for using different interpolating functions, among them
Gaussian process regression. The results are all compatible and the differences
decrease with the increased number of simulations. We also tested the
consistency of the results by dropping randomly 10\% of the simulations 
used to produce the interpolated maps.
Note that we have restricted the grids to spin magnitudes $\leq0.85$ in order
to produce interpolation maps and avoid at this stage extrapolations to
larger spins until we produce enough simulations in the $\geq0.85$ region.

\begin{figure}
\includegraphics[angle=0, width=0.85\columnwidth]{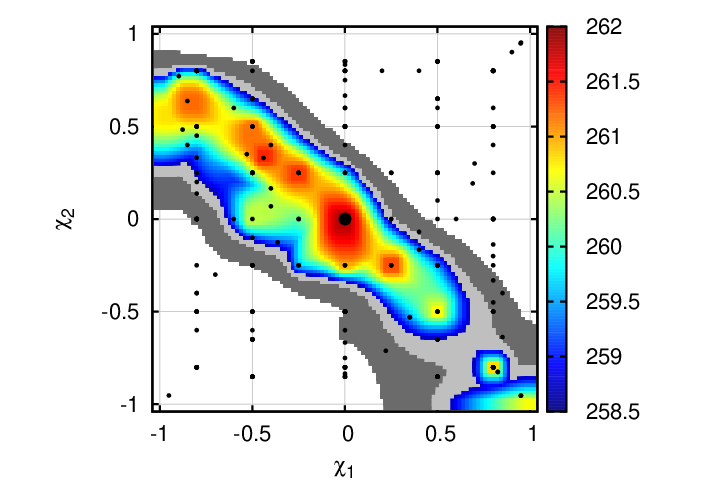}\\
\includegraphics[angle=0, width=0.85\columnwidth]{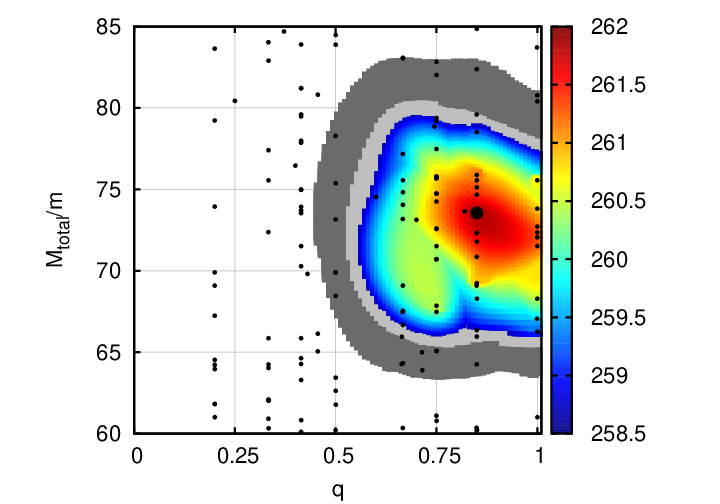}\\
\includegraphics[angle=0, width=0.85\columnwidth]{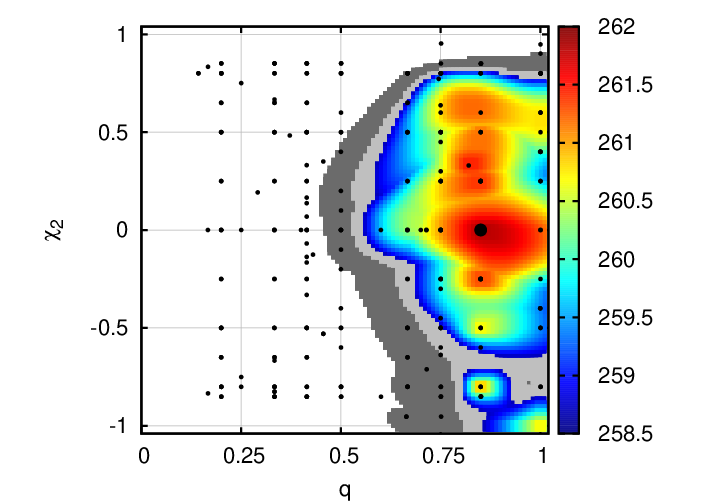}\\
\includegraphics[angle=0, width=0.85\columnwidth]{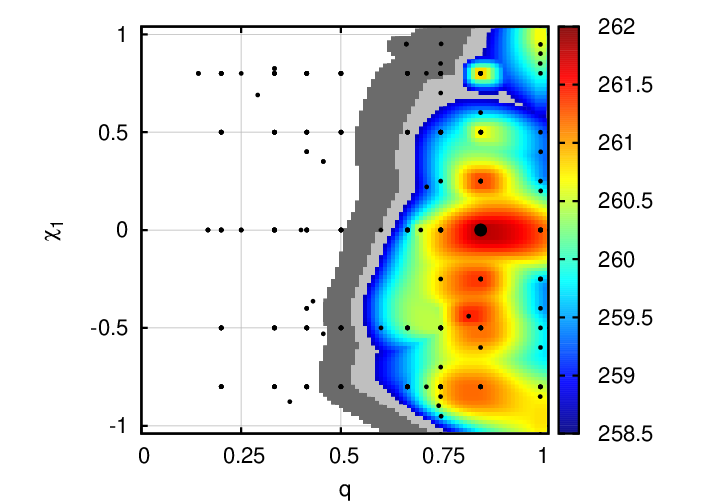}
  \caption{90\% confidence interval heat maps of the GW150914 likelihood for the aligned binary
mass ratio and individual spin parameters.  The dark grey region
 constitutes the 99.7\% ($3\sigma$) confidence interval range, 
 and the light grey is the 95\% ($2\sigma$) range. The colored region shows the
 $\ln {\cal L}$ of the values within the 90\% confidence interval. 
The black points indicate the placement of the numerical simulations.
\label{fig:errors}}
\end{figure}

Fig.~\ref{fig:errors} displays the error estimates of the aligned spins and
mass ratio binary parameters for GW150914 at 90\%, 95\% ($2\sigma$), and 99.7\% ($3\sigma$) confidence levels. Note that the diagonal shape of
higher likelihoods the first panel is in part inherited by the symmetry of the $q=1$ case, given that the binary has comparable masses. The elongated (with bubbles) diagonal shape of the first panel translates into a vertical shapes in the last two panels.

The $90\%$ confidence level gives 
\begin{align*}
0.570  & < q\ \ \ \,<1.00,\\
0.00 & < |\chi_1| <1.00,\\
0.00 & < |\chi_2| <0.78,\\
-0.44& < \chi_{\mathrm{eff}} < 0.14,\\
-0.44& < S_{hu} < 0.14,\\
66.3 & < M_{total} < 79.2
\end{align*} 
Where $M_{total}$ is given in solar mass $M_\odot$ units.

Compare these values to the GW150914 properties paper \cite{TheLIGOScientific:2016wfe} 
\begin{align*}
0.62 & <q\ \ \ \,<      0.99,\\
0.04 & <|\chi_1| <0.90,\\
0.03 & <|\chi_2| <0.78,\\
-0.29& < \chi_{\mathrm{eff}} < 0.1,\\
66.1 & < M_{total} < 75.2
\end{align*}

\begin{figure}
\includegraphics[angle=0, width=\columnwidth]{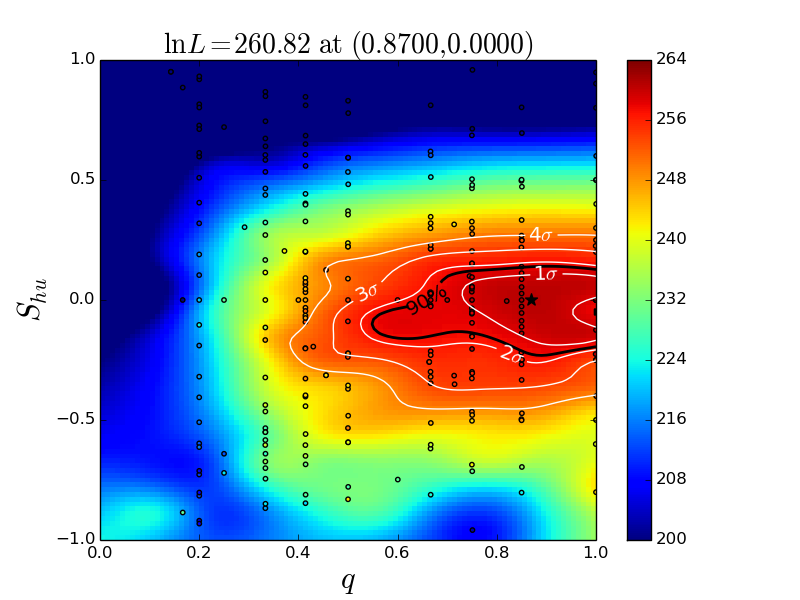}\\
\includegraphics[angle=0, width=\columnwidth]{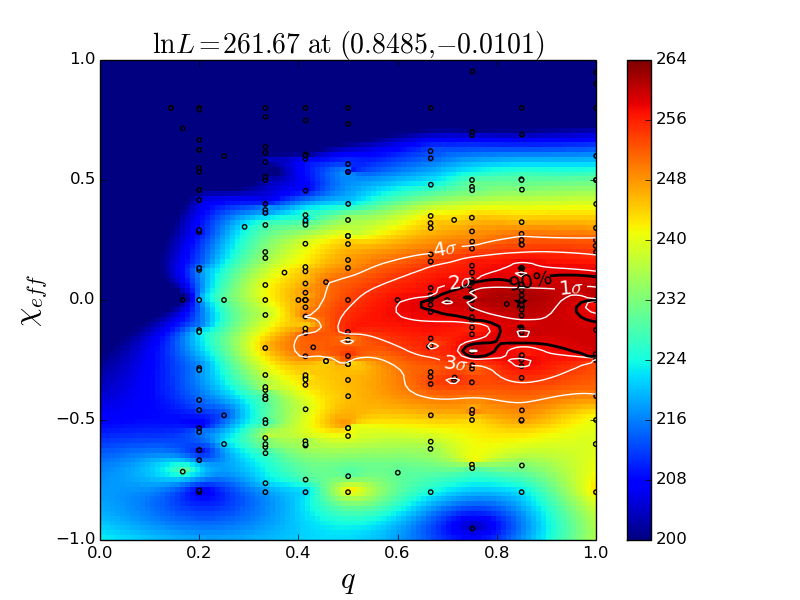}
  \caption{Heat maps of the GW150914 likelihood for the aligned binary
    with effective variables $S_{hu}$ and $\chi_{eff}$ versus mass ratios
    using linear interpolation.
    In black the 90\% confidence contours and the interpolated
    $\ln {\cal L}$ maximum is given in each panel's title and denoted by the *
    in the plots.
\label{fig:Shu}}
\end{figure}

Fig.~\ref{fig:Shu} displays a comparative analysis of the single spin approximations to aligned binaries using a linear interpolation. The upper panel presents our preferred variables for
the spin, $S_{hu}$ 
\beq
m^2\,S_{hu}=\left((1+\frac{1}{2q})\,\vec{S}_1+(1+\frac{1}{2}q)\,\vec{S}_2\right)\cdot\hat{L},
\eeq
to describe the leading effect of hangup on
the waveforms \cite{Healy:2018swt}. 
The lower panel displays a comparative heatmap using
the common approximate model variable \cite{Ajith:2009bn} 
$$m^2\,\chi_{eff}=\left((1+\frac1q)\,\vec{S}_1+(1+q)\,\vec{S}_2\right)\cdot\hat{L}.$$
The latter exhibits some ``pinch'' points around some simulations suggesting a remaining degeneracy by using $\chi_{eff}$.
Such features are not seen using the (normalized) variable
${S}_{hu}$, 
which represents a better fitting to waveform phases as shown in
\cite{Healy:2018swt}, suggesting again that it is a better (or at least
a valid alternative) choice to describe aligned binaries.

\begin{figure}
\includegraphics[angle=0,width=0.7\columnwidth]{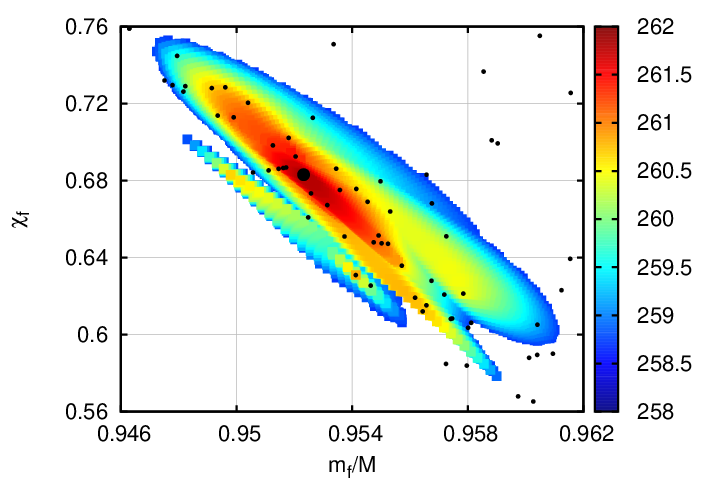}\\
\includegraphics[angle=0,width=0.7\columnwidth]{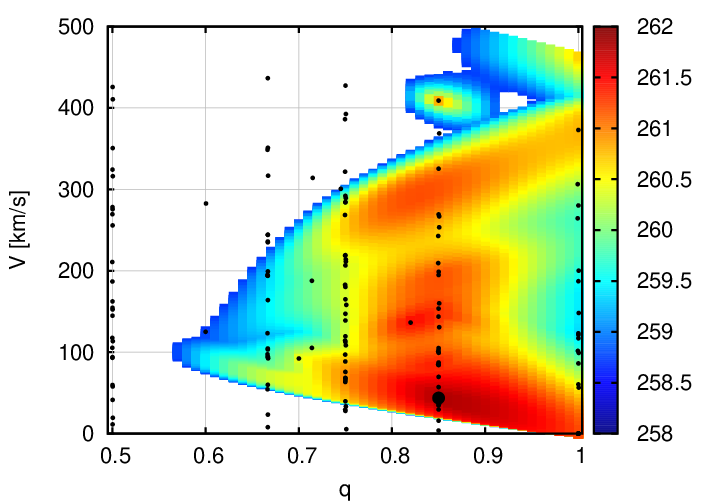}\\
\includegraphics[angle=0,width=0.7\columnwidth]{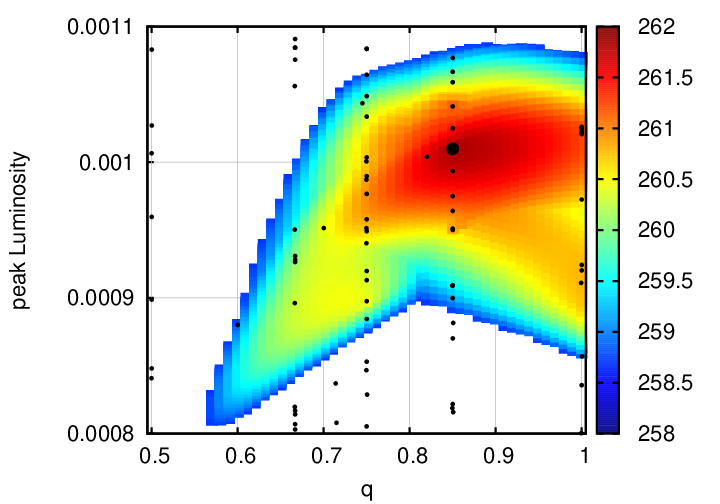}\\
\includegraphics[angle=0,width=0.7\columnwidth]{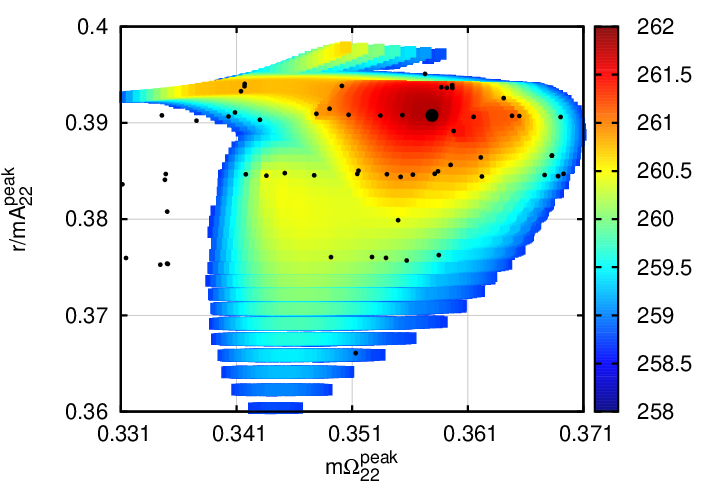}
\caption{Final parameter space heatmaps for simulations that fall within 
the 90\% confidence interval for the final mass, spin, recoil, peak luminosity, and 
orbital frequency and strain amplitude at peak strain.
A maximum $\ln {\cal L}$ is reached for $m_f/m=0.952$, $\chi_f=0.683$, $V=44$ km/s,
$L^{peak}=1.01e-3$, $m\Omega_{22}^{peak}=0.358$, and $(r/m)A^{peak}_{22}=0.391$.
\label{fig:remnant90CI}}
\end{figure}

The 3D interpolated results give a maximum $\ln {\cal L}$ of 261.8 at 
$(M_{total}, q, \chi_1, \chi_2) = ( 73.6, 0.8500, 0.0000, 0.0000 )$.  
Values of the final mass and spin 
for this point are $0.952$ and $0.683$, respectively, and the recoil velocity is $44$ km/s.
The mean values from the GW150914 properties paper~\cite{TheLIGOScientific:2016wfe} are
0.955 and 0.67 for the final mass and spin, respectively.
Converting the final mass to energy radiated and calculating the ranges in these final
parameters from the simulations that fall within the 90\% confidence interval
as shown in Fig.~\ref{fig:remnant90CI}, we find 
\begin{align*}
0.039 & < E_{rad}/m < 0.053\\
0.578 & < \chi_f < 0.753\\
0 & < V_{recoil} < 492\text{[km/s]}
\end{align*}
Comparing these ranges to the GW150914 properties paper \cite{TheLIGOScientific:2016wfe} 
(and converting from total mass and final mass to energy radiated and 
propagating the errors appropriately) 
\begin{align*}
0.041 & < E_{rad}/m < 0.049 \\
0.60 & < \chi_f < 0.72
\end{align*}
Note that a final recoil velocity is not estimated in
\cite{TheLIGOScientific:2016wfe}.


\subsection{Precessing binaries}

\begin{figure*}
\includegraphics[angle=0, width=\columnwidth]{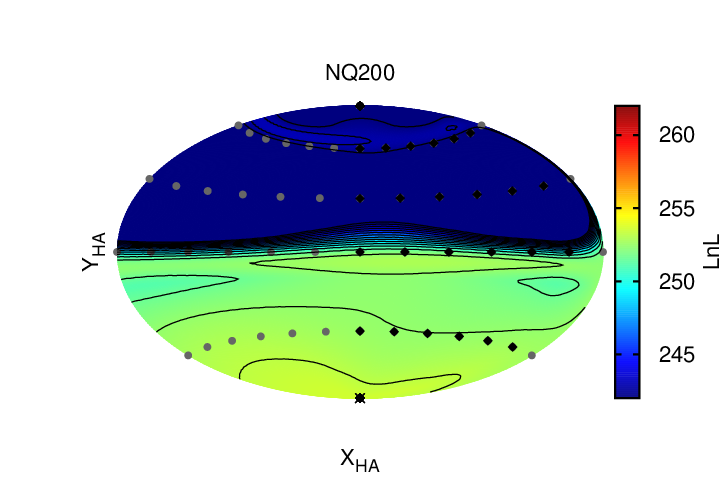}
\includegraphics[angle=0, width=\columnwidth]{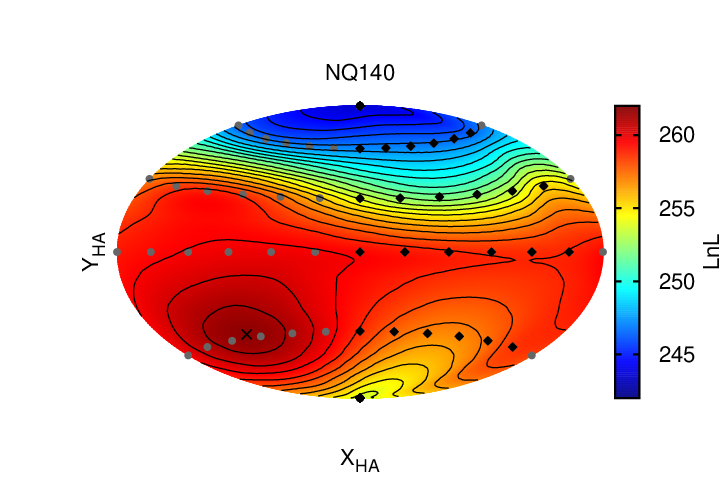}\\
\includegraphics[angle=0, width=\columnwidth]{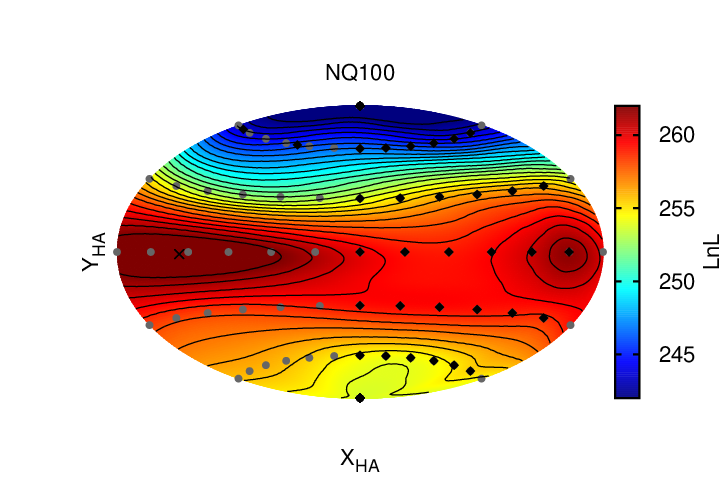}
\includegraphics[angle=0, width=\columnwidth]{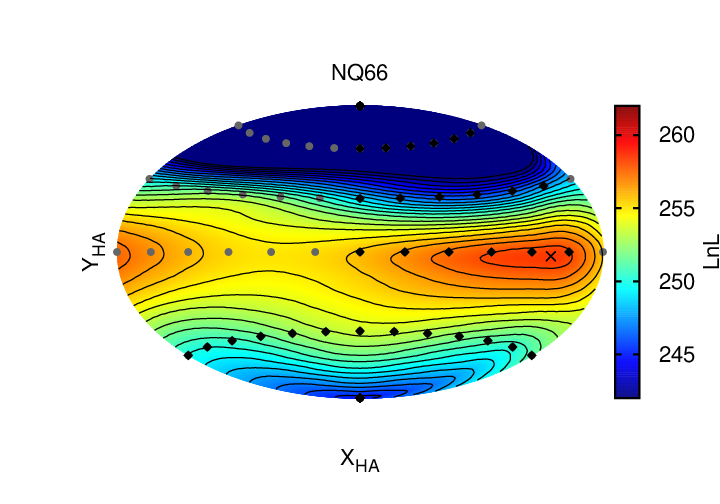}\\
\includegraphics[angle=0, width=\columnwidth]{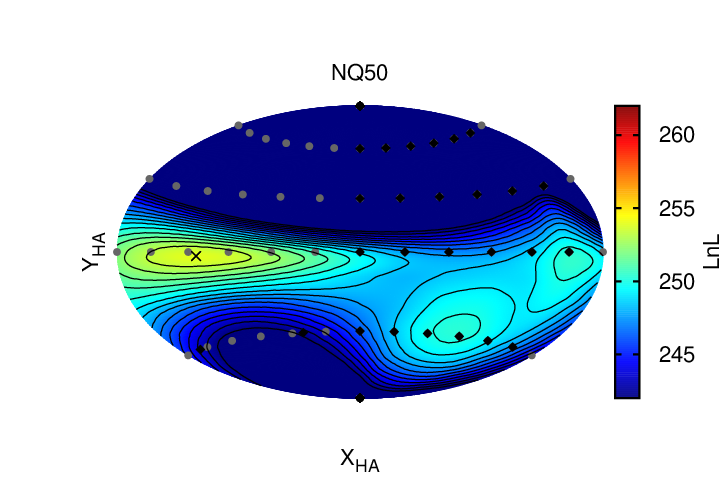}
\includegraphics[angle=0, width=\columnwidth]{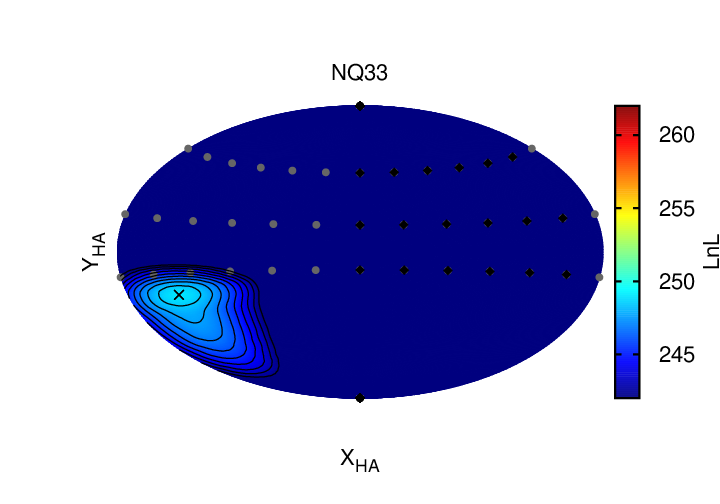}
\caption{Heat maps of the GW150914 likelihood for each of the six mass ratio panels covering form $q=2$ to $q=1/3$ (labeled from NQ200 to NQ33 respectively)
  and large black hole spin oriented over the sphere
(interpolated using multiquadric radial basis functions between simulations).
The individual panel with $q=1$ contains the highest likelihood (near the orbital plane orientation), and it is bracketed by the $q=1.4$ and $q=0.66$ panels
($q>1$ here means the smaller black hole is the one spinning).
We have used Hammer-Aitoff coordinates
$X_{HA}, Y_{HA}$, to represent the map and level curves. 
The interpolated $\ln {\cal L}$ maximum location is denoted by the an x
    in the plots, the black points are simulations, and the gray points are extrapolated
    simulations using the sinusoidal dependence of the azimuthal angle.
    \label{fig:heatmapprec}}
\end{figure*}

An analogous study of GW150914 using the aligned spin binaries above can be done in a completely independent way with a set of precessing binaries. We supplement the new simulations in this catalog release with those reported in Ref.~ \cite{Lousto:2012gt,Zlochower:2015wga} for mass-ratio families of binaries with 
one spinning black hole pointing along 32 different orientations.
The results of evaluation of the $\ln {\cal L}$ for a set of six different
mass ratio families are displayed in Fig.~\ref{fig:heatmapprec}.
We find that the highest likelihood,is displayed in the $q=1$ panel
and spin orientation near the equatorial (orbital) plane. This result
is consistent with the low $S_{hu}$ (or low $\chi_{eff}$) displayed
in Fig.~\ref{fig:Shu}. 
The $q=1$ highest likelihood is bracketed by the $q=1.4$ and $q=0.66$ panels with the former having a larger $\ln {\cal L}$  than the later, indicating that the
optimal configuration should have a
mass ratio between $q=1$ and $q=1.40$ (that corresponds to a case of $q=1/1.4=0.714$ with the smaller black hole spinning). This again is in agreement with
the previous analysis involving only aligned simulations indicating a preference for mass ratios around $q=0.85$. 

Let us note that this precessing simulations analysis is completely independent from the previous nonprecessing, aligned spins, and they do not share simulations in common and yet lead to a similar range of parameters (for the most robust mass ratios and projections of the spins onto the orbital angular momentum).
While the analysis would benefit from more simulations to populate this 4D parameter space (and this will be one of the subjects of a new catalog release), it is encouraging that consistent results are already found with this minimal set of nearly 200 simulations. we populated each panel with 4 sets
of initial $\theta=30,60,90,135$ degrees orientations (note that new configurations will supplant $\theta=135$ for 120 and 150 degrees) 
and six $\phi=0,30,60,90,120,150$ plus the poles and a few control simulations on the $\phi<0$ degrees (See Fig.~7 of Ref.~\cite{Healy:2017abq}). 
To extrapolate into the western hemisphere, we exploit the symmetry of the parameter space and fit a sinusoidal function to the available numerical simulations.
Instead of plotting in the angles $\theta$ and $\phi$, we plot in the Hammer-Aitoff coordinates
\footnote{Weinstein, Eric W.``Hammer-Aitoff Equal-Area Projection.'' From MathWorld--A Wolfram Web Resource. http://mathworld.wolfram.com/Hammer-AitoffEqual-AreaProjection.html}, which is a
coordinate system where the whole angular space can be viewed as a 2D map.  The points at the top left and bottom left
are the poles, $\theta=0$ at the top, and $\theta=\pi$ at the bottom.  The line connecting the two is the $\phi=0$ line.  As
you move from left to right from the center, $\phi$ increases from 0 to 180 degrees, and from right to left, $\phi$ decreases from 0 to -180 degrees.

\subsection{Estimation of extrinsic parameters}

\begin{figure}
\includegraphics[angle=0, width=0.95\columnwidth]{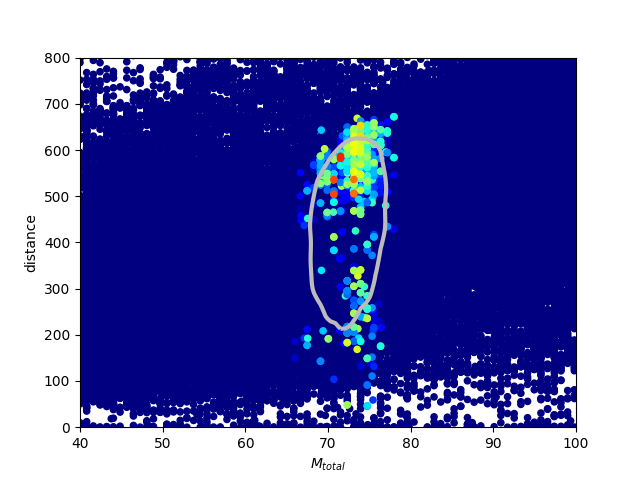}
\includegraphics[angle=0, width=0.95\columnwidth]{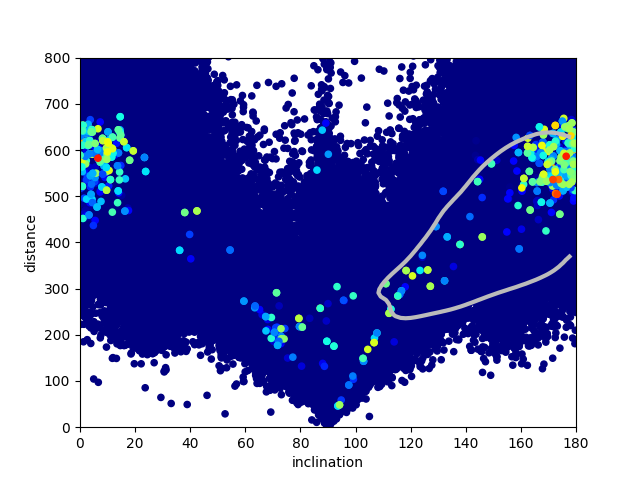}
\includegraphics[angle=0, width=\columnwidth]{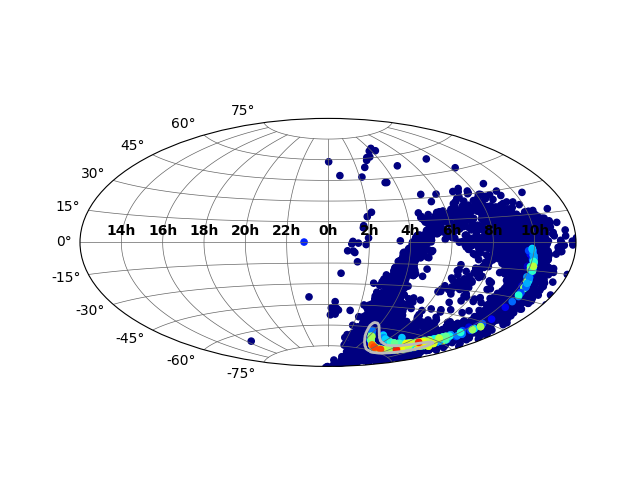}
  \caption{We use the results of the Monte-Carlo intrinsic log-likelihood calculations
    (100 samples in $M_{total}$ for each simulation in the catalog) to estimate the extrinsic
    parameters of GW150914.  The gray boundary denotes the public LIGO GWTC-1 data
    and the colored points indicate simulations which fell within the 
    $\ln{\cal L} > \mathrm{max}\ln{\cal L} - 3.125$, or roughly the 90\% confidence
    interval. The dark blue background points denote simulations outside of the 90\% confidence
    interval. 
\label{fig:extrinsic}}
\end{figure}

To complete the parameter determination from only the numerical relativity
simulations we proceed to compute the distribution of the sky location,
distance and binary orientation from the evaluations of the $\ln {\cal L}$
for each of our simulations on a total mass grid of 100 points between 
40 and 120 $M_{\odot}$. The results for GW150914 are displayed
in Fig. \ref{fig:extrinsic}.  The gray boundaries are calculated from
the LIGO GWTC-1 public data \footnote{https://dcc.ligo.org/LIGO-P1800370/public}
for GW150914 by first constructing a 2D kernel destiny estimation and using 
numpy's percentile function.  Simulations within a $\ln {\cal L}$ cut of 3.125
of the max are included in the 90\% CI interval.  As with the intrinsic parameters,
the results are consistent but less confined than the LIGO results.
Our results seem also to be compatible with the original localization in sky estimates displayed in Ref.~\cite{Abbott:2016gcq}

\subsection{Simulated versus signal waveform comparison}

\begin{figure*}
\includegraphics[angle=0, width=1.8\columnwidth]{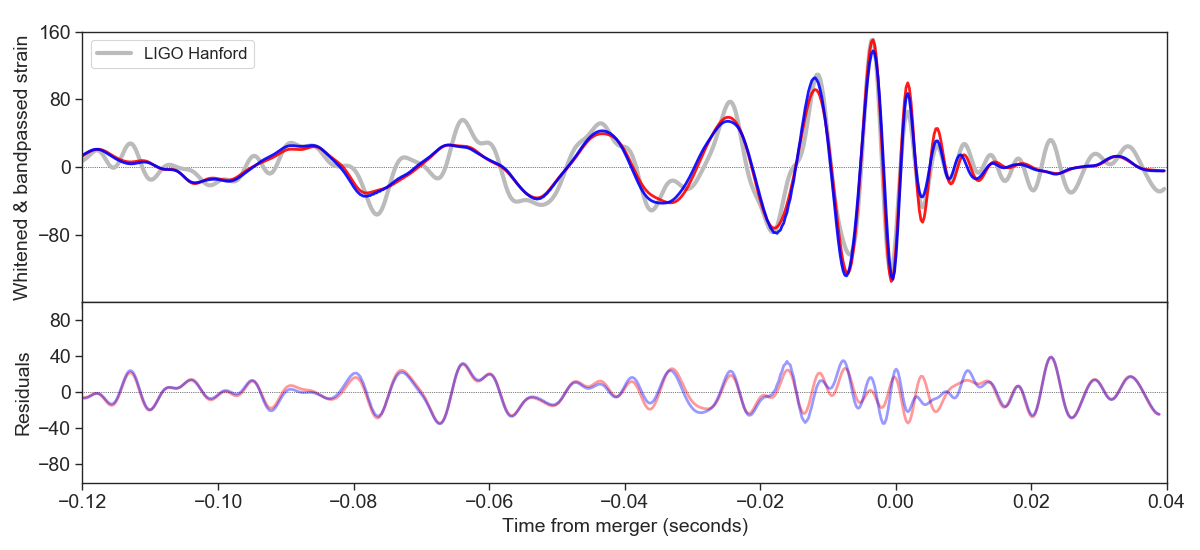}\\
\includegraphics[angle=0, width=1.8\columnwidth]{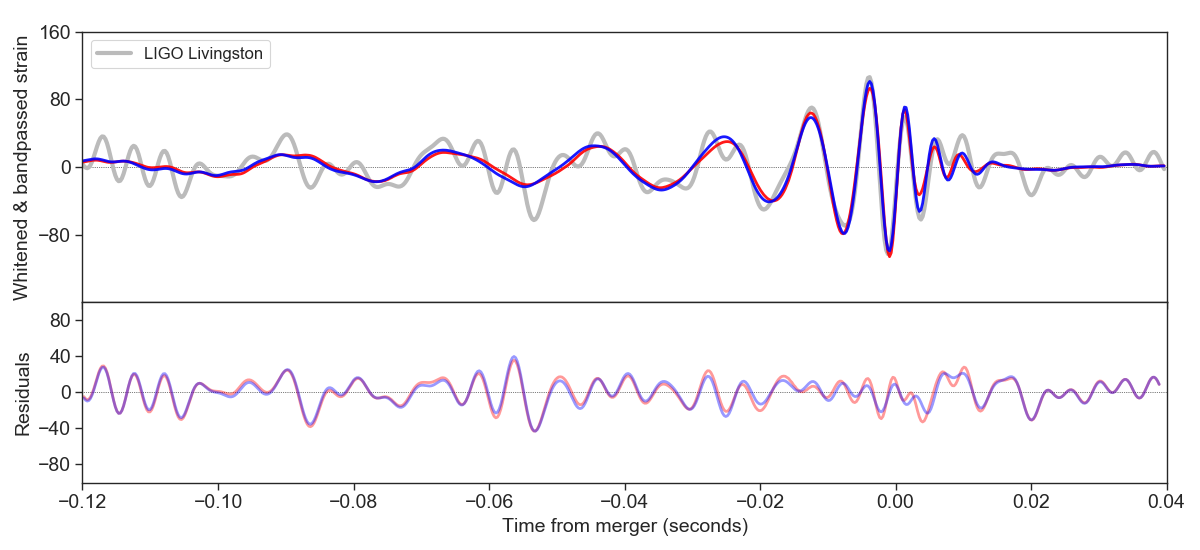}
  \caption{ Direct comparison of the highest $\ln{\cal L}$ nonprecessing simulation 
            (RIT:BBH:0113 in red) and precessing simulation (RIT:BBH:0126 in blue) to the 
            Hanford (top) and Livingston (bottom) GW150914 signals.  The bottom panel
            in each figure shows the residual between the whitened NR waveform and detector
            signal.
\label{fig:simstrain}}
\end{figure*}

We use standard techniques \cite{Abbott:2016apu,Lange:2017wki} to directly compare GW150914 to our simulations.
For each simulation, direct comparison of our simulations to the
data selects a fiducial total mass which best fits the
observations, as measured by the marginalized likelihood.
We can for each simulation select the binary extrinsic parameters
like event time and sky location which maximize the likelihood of the data,
given our simulation and mass. Then, using these extrinsic parameters,
we evaluate the expected detector response in the LIGO Hanford (H1)
and Livingston (L1) instruments. Figure \ref{fig:simstrain} displays
these reconstructions for the highest log-likelihood NR waveform of the
nonprecessing and precessing simulations labeled as
RIT:BBH:0113 and RIT:BBH:0126 in our catalog.  The details of these 
simulations are provided in Table~\ref{tab:gwcomp}.
They directly compare to the signals as observed by LIGO H1 and L1
and with each other. The lower panel shows the residuals of the
signals with respect to the RIT simulations.
A similar analysis was performed in Ref.~\cite{Healy:2017abq}, Figures 4-6,
for the GW170104 event.

\begin{table*}
\caption{Highest $\ln{\cal L}$ nonprecessing and precessing simulations.  
The nonprecessing simulation has highest overall $\ln{\cal L}$, and
the precessing simulation has 13th highest.
\label{tab:gwcomp}}
\begin{ruledtabular}
\begin{tabular}{lcccccc}
Config. & $q$ & $\vec \chi_1$ & $\vec \chi_2$ & $\vec S_{hu}/m^2$ & $M_{total}/M_{\odot}$ & $\ln{\cal L}$ \\
\hline
RIT:BBH:0113 & 0.85 & (0, 0, 0)             & (0, 0, 0)           & (0, 0, 0)             & 73.6 & 261.8\\
RIT:BBH:0126 & 0.75 & (-0.46, -0.48, -0.44) & (0.06, -0.38, 0.12) & (-0.15, -0.42, -0.11) & 72.5 & 260.5
\end{tabular}
\end{ruledtabular}
\end{table*}


\section{Conclusions and Discussion}\label{sec:Discussion}

The
breakthroughs~\cite{Pretorius:2005gq,Campanelli:2005dd,Baker:2005vv}
in numerical relativity were instrumental in identifying the first
detection of gravitational waves \cite{TheLIGOScientific:2016wfe} with
the merger of two black holes.  We have shown in this paper that the
use of numerical relativity waveform catalogs provides a consistent
method for parameter estimation from the observed gravitational waves
from merging binary black holes (See also Refs.
\cite{Abbott:2016apu,Lange:2017wki,Kumar:2018hml}.
It is worthwhile stressing here that
this current method of direct comparison of the gravitational wave
signal with numerical waveforms does not rely {\it at all} on any 
information from
the phenomenological models \cite{Babak:2016tgq,Hannam:2013oca} (Phenom
or SEOBNR).
It also shows that with the current aligned spin coverage one can
successfully carry out parameter estimations with results,
at least as good as with the phenomenological models
\cite{TheLIGOScientific:2016wfe}; see Lange et al (in prep).
We also note that any new simulation produced (for instance targeted to
followup any new detection or catalog expansions) will contribute to
improve the binary parameter coverage, thus reducing the interpolation
error. The next step will be
reduce the extrapolation error
at very high spins by adding more simulations with spin magnitudes 
above 0.90. Also the extension of the family of simulations displayed
in Fig. \ref{fig:heatmapprec} to smaller mass ratios, i.e. $q<1/3$. Coverage for
low total binary masses (below $20M_\odot$), in turn, would require longer
full numerical simulations or hybridization of the current NR waveforms
with post-Newtonian waveforms.

The next area of development for the numerical relativity waveform
catalogs is the coverage of precessing binaries. Those require
expansions of the parameter space to seven dimensions (assuming
negligible eccentricity), and is being carried out in a hierarchical
approach by neglecting the effects of the spin of the secondary black
holes, which is a good assumption for small mass ratios. This approach
has proven also successful when applied to
GW170104~\cite{Healy:2017abq}.  It required an homogeneous set of
simulations since the differences in $\ln {\cal L}$ are subtle.
The comparison of
different approaches to solve the binary black hole problem has
produced an excellent agreement for the GW150914
\cite{Lovelace:2016uwp} and GW170104~\cite{Healy:2017abq}, including
higher (up to $\ell=5$) modes.  This leads to the possibility of regularly
using multiple catalogs of numerical relativity waveforms to further improve
parameter coverage as started in \cite{Abbott:2016apu}.
In a follow up paper we plan to use this upgraded catalog to evaluate
the parameters of the ten binary black hole mergers reported recently
by LIGO-Virgo O1-O2 observing runs\cite{LIGOScientific:2018jsj}.

Aside from the interest in producing waveforms for direct comparison
with observation, the simulations of orbiting black-hole binaries
produce information about the final remnant of the merger of the two
holes.  Numerous empirical formulas relating the initial parameters
$(q,\vec\chi_1,\vec\chi_2)$ (individual masses and spins) of the
binary to those of the final remnant $(m_f,\vec\chi_f,\vec{V}_f)$ have
been proposed. These include formulas for the final mass, spin, and
recoil velocity
\cite{Barausse:2012qz,Rezzolla:2007rz,Hofmann:2016yih,Jimenez-Forteza:2016oae,Lousto:2009mf,Lousto:2013wta,Hemberger:2013hsa,Healy:2014yta,Zlochower:2015wga},
the computation of the peak frequency of the (2,2) mode
$\Omega_{22}^{peak}$, peak waveform
amplitude $A_{22}^{peak}$
~\cite{Healy:2017mvh,Healy:2018swt} and peak luminosity
\cite{TheLIGOScientific:2016wfe,TheLIGOScientific:2016pea,Healy:2016lce,Keitel:2016krm}.
Those formulas in turn provide further tools to extract information
from the observation of gravitational waves, see for instance our
Fig.~\ref{fig:remnant90CI}.


\begin{acknowledgments}

The authors thank A. Williamson for assistance with Fig.~\ref{fig:simstrain}.
The authors also gratefully acknowledge the National Science Foundation
(NSF) for financial support from Grants No.\ PHY-1607520,
No.\ PHY-1707946, No.\ ACI-1550436, No.\ AST-1516150,
No.\ ACI-1516125, No.\ PHY-1726215.  This work used the Extreme
Science and Engineering Discovery Environment (XSEDE) [allocation
  TG-PHY060027N], which is supported by NSF grant No. ACI-1548562.
Computational resources were also provided by the NewHorizons, BlueSky
Clusters, and Green Prairies at the Rochester Institute of Technology,
which were supported by NSF grants No.\ PHY-0722703, No.\ DMS-0820923,
No.\ AST-1028087, No.\ PHY-1229173, and No.\ PHY-1726215.
\end{acknowledgments}


\appendix

\section{Tables of initial data and results of the new simulations}\label{app:ID}

In this appendix we provide tables with the relevant BBH
configuration details.  In Table \ref{tab:ID}, we provide the
initial data parameters  used to start
the full numerical evolutions. In Tables \ref{tab:IDr} and \ref{tab:IDr_prec},
we provide the binary mass and spin parameters after 
they settle into a more physical value after radiating and absorbing
the spurious gravitation wave content from the initial mathematical 
choice of conformal flatness.  These relaxed values are calculated at  
a fiducial $t=200M$.

In Table \ref{tab:ecc} we provide the initial orbital frequency
and eccentricity, as well as the number of orbits to merger and the
final eccentricity. The eccentricity is expected to be reduced from its 
initial value by gravitational radiation, at a rate proportional to
$d^{19/12}$ according to \cite{Peters:1964zz}, with $d$, the separation of the
binary (see, for instance, Fig.~6 of Ref.~\cite{Mroue:2010re} or
Fig.~9 in Ref.~\cite{Lousto:2015uwa}).

Finally, In Table \ref{tab:spinerad}, we provide
the values of the energy radiated during the simulation
and the final black hole spin as measured through the (most accurate)
isolated horizon formalism \cite{Dreyer02a}.

\begin{longtable*}{lccccccccccccc}
\caption{Initial data parameters for the quasi-circular
configurations with a smaller mass black hole (labeled 1),
and a larger mass spinning black hole (labeled 2). The punctures are located
at $\vec r_1 = (x_1,0,0)$ and $\vec r_2 = (x_2,0,0)$, with individual linear momenta
$P=\pm (P_r, P_t,0)$, spin magnitudes $|S_i|$, puncture mass parameters
$m^p/m$, horizon (Christodoulou) masses $m^H/m$, total ADM mass
$M_{\rm ADM}$, and dimensionless spins $|a/m_H| = |S/m_H^2|$.
The spin directions for the precessing simulations are given in
the catalog and in Tab.~\ref{tab:IDr_prec}.}
\label{tab:ID}\\
\hline
\hline
Run   & $x_1/m$ & $x_2/m$  & $P_r/m$    & $P_t/m$ & $m^p_1/m$ & $m^p_2/m$ & $|S_1/m^2|$ & $|S_2/m^2|$ & $m^H_1/m$ & $m^H_2/m$ & $M_{\rm ADM}/m$ & $|a_1/m_1^H|$ & $|a_2/m_2^H|$\\
\hline
\endfirsthead

\multicolumn{14}{c}
{{\tablename\ \thetable{} -- continued from previous
page}}\\
\hline
Run   & $x_1/m$ & $x_2/m$  & $P_r/m$    & $P_t/m$ & $m^p_1/m$ & $m^p_2/m$ & $|S_1/m^2|$ & $|S_2/m^2|$ & $m^H_1/m$ & $m^H_2/m$ & $M_{\rm ADM}/m$ & $|a_1/m_1^H|$ & $|a_2/m_2^H|$\\
\hline
\endhead
\hline \multicolumn{14}{r}{{Continued on next page}} \\
\hline
\endfoot
\hline\hline
\endlastfoot

RIT:BBH:0127 & -5.71 & 4.29 & 0 & 0.09386 & 0.2578 & 0.347 & 0.1469 & 0.2612 & 0.4286 & 0.5714 & 0.9901 & 0.8 & 0.8 \\
RIT:BBH:0128 & -6.67 & 3.33 & 0 & 0.08542 & 0.1992 & 0.4072 & 0.08889 & 0.3556 & 0.3333 & 0.6667 & 0.9911 & 0.8 & 0.8 \\
RIT:BBH:0129 & -7.50 & 2.50 & 0 & 0.07226 & 0.1485 & 0.4604 & 0.05 & 0.45 & 0.25 & 0.75 & 0.9925 & 0.8 & 0.8 \\
RIT:BBH:0130 & -5.71 & 4.29 & 0 & 0.09362 & 0.2578 & 0.3469 & 0.1469 & 0.2612 & 0.4286 & 0.5714 & 0.9901 & 0.8 & 0.8 \\
RIT:BBH:0131 & -6.67 & 3.33 & 0 & 0.08502 & 0.1992 & 0.4072 & 0.08889 & 0.3556 & 0.3333 & 0.6667 & 0.9911 & 0.8 & 0.8 \\
RIT:BBH:0132 & -7.50 & 2.50 & 0 & 0.07189 & 0.1485 & 0.4603 & 0.05 & 0.45 & 0.25 & 0.75 & 0.9925 & 0.8 & 0.8 \\
RIT:BBH:0136 & -10.00 & 5.00 & 0 & 0.06464 & 0.3248 & 0.546 & 0 & 0.2667 & 0.3333 & 0.6667 & 0.9934 & 0 & 0.6 \\
RIT:BBH:0220 & -7.50 & 5.00 & -4.96e-04 & 0.08177 & 0.3889 & 0.3666 & 0 & 0.288 & 0.4 & 0.6 & 0.9921 & 0 & 0.8 \\
RIT:BBH:0221 & -10.83 & 2.17 & -1.51e-04 & 0.04638 & 0.09955 & 0.7352 & 0.02222 & 0.3472 & 0.1667 & 0.8333 & 0.9957 & 0.8 & 0.5 \\
RIT:BBH:0222 & -10.83 & 2.17 & -1.20e-04 & 0.04418 & 0.09964 & 0.7352 & 0.02222 & 0.3472 & 0.1667 & 0.8333 & 0.9954 & 0.8 & 0.5 \\
RIT:BBH:0223 & -7.03 & 5.97 & -4.30e-04 & 0.08115 & 0.279 & 0.5162 & 0.1689 & 0.07305 & 0.4595 & 0.5405 & 0.992 & 0.8 & 0.25 \\
RIT:BBH:0224 & -10.83 & 2.17 & -1.09e-04 & 0.04312 & 0.09967 & 0.515 & 0.02222 & 0.5556 & 0.1667 & 0.8333 & 0.9954 & 0.8 & 0.8 \\
RIT:BBH:0226 & -10.83 & 2.17 & -1.27e-04 & 0.0447 & 0.09961 & 0.8278 & 0.02222 & 0 & 0.1667 & 0.8333 & 0.9955 & 0.8 & 0 \\
RIT:BBH:0227 & -7.03 & 5.97 & -3.89e-04 & 0.07931 & 0.279 & 0.5163 & 0.1689 & 0.07305 & 0.4595 & 0.5405 & 0.9918 & 0.8 & 0.25 \\
RIT:BBH:0228 & -6.50 & 6.50 & -3.94e-04 & 0.07983 & 0.2579 & 0.4557 & 0.2125 & 0.1 & 0.5 & 0.5 & 0.9918 & 0.85 & 0.4 \\
RIT:BBH:0230 & -7.50 & 5.00 & -4.73e-04 & 0.08102 & 0.3889 & 0.3666 & 0 & 0.288 & 0.4 & 0.6 & 0.992 & 0 & 0.8 \\
RIT:BBH:0231 & -7.50 & 5.00 & -4.74e-04 & 0.08104 & 0.3889 & 0.3666 & 0 & 0.288 & 0.4 & 0.6 & 0.992 & 0 & 0.8 \\
RIT:BBH:0232 & -7.03 & 5.97 & -5.02e-04 & 0.08372 & 0.2789 & 0.3295 & 0.1689 & 0.2337 & 0.4595 & 0.5405 & 0.9924 & 0.8 & 0.8 \\
RIT:BBH:0233 & -7.50 & 5.00 & -4.76e-04 & 0.0811 & 0.3889 & 0.3666 & 0 & 0.288 & 0.4 & 0.6 & 0.9921 & 0 & 0.8 \\
RIT:BBH:0234 & -7.50 & 5.00 & -4.77e-04 & 0.08113 & 0.3889 & 0.3666 & 0 & 0.288 & 0.4 & 0.6 & 0.9921 & 0 & 0.8 \\
RIT:BBH:0235 & -7.50 & 5.00 & -4.76e-04 & 0.0811 & 0.3889 & 0.3666 & 0 & 0.288 & 0.4 & 0.6 & 0.9921 & 0 & 0.8 \\
RIT:BBH:0236 & -7.50 & 5.00 & -4.74e-04 & 0.08104 & 0.3889 & 0.3666 & 0 & 0.288 & 0.4 & 0.6 & 0.992 & 0 & 0.8 \\
RIT:BBH:0237 & -6.86 & 5.14 & -5.90e-04 & 0.08569 & 0.311 & 0.5391 & 0.1286 & 0.09796 & 0.4286 & 0.5714 & 0.9917 & 0.7 & 0.3 \\
RIT:BBH:0238 & -6.86 & 5.14 & -6.35e-04 & 0.08695 & 0.3701 & 0.2953 & 0.09184 & 0.2776 & 0.4286 & 0.5714 & 0.9919 & 0.5 & 0.85 \\
RIT:BBH:0239 & -6.86 & 5.14 & -5.61e-04 & 0.08478 & 0.3702 & 0.2953 & 0.09184 & 0.2776 & 0.4286 & 0.5714 & 0.9916 & 0.5 & 0.85 \\
RIT:BBH:0240 & -7.03 & 5.97 & -4.87e-04 & 0.08322 & 0.3983 & 0.2793 & 0.1056 & 0.2484 & 0.4595 & 0.5405 & 0.9922 & 0.5 & 0.85 \\
RIT:BBH:0241 & -5.62 & 4.19 & -9.95e-04 & 0.09488 & 0.1605 & 0.3708 & 0.1634 & 0.2534 & 0.4271 & 0.5729 & 0.9901 & 0.8957 & 0.7719 \\
RIT:BBH:0242 & -7.50 & 5.00 & -4.74e-04 & 0.08104 & 0.3889 & 0.3666 & 0 & 0.288 & 0.4 & 0.6 & 0.992 & 0 & 0.8 \\
RIT:BBH:0243 & -7.50 & 5.00 & -4.76e-04 & 0.0811 & 0.3889 & 0.3666 & 0 & 0.288 & 0.4 & 0.6 & 0.9921 & 0 & 0.8 \\
RIT:BBH:0244 & -7.50 & 5.00 & -4.74e-04 & 0.08104 & 0.3889 & 0.3666 & 0 & 0.288 & 0.4 & 0.6 & 0.992 & 0 & 0.8 \\
RIT:BBH:0245 & -8.67 & 4.33 & -3.67e-04 & 0.0735 & 0.2871 & 0.3467 & 0.05556 & 0.3778 & 0.3333 & 0.6667 & 0.9929 & 0.5 & 0.85 \\
RIT:BBH:0246 & -10.83 & 2.17 & -1.34e-04 & 0.04524 & 0.0996 & 0.8277 & 0.02222 & 0 & 0.1667 & 0.8333 & 0.9955 & 0.8 & 0 \\
RIT:BBH:0247 & -8.67 & 4.33 & -2.97e-04 & 0.07024 & 0.2873 & 0.3467 & 0.05556 & 0.3778 & 0.3333 & 0.6667 & 0.9926 & 0.5 & 0.85 \\
RIT:BBH:0248 & -7.50 & 5.00 & -4.73e-04 & 0.08102 & 0.3889 & 0.3666 & 0 & 0.288 & 0.4 & 0.6 & 0.992 & 0 & 0.8 \\
RIT:BBH:0249 & -7.03 & 5.97 & -3.85e-04 & 0.07913 & 0.3718 & 0.5299 & 0.1267 & 0 & 0.4595 & 0.5405 & 0.9917 & 0.6 & 0 \\
RIT:BBH:0250 & -7.50 & 5.00 & -4.77e-04 & 0.08113 & 0.3889 & 0.3666 & 0 & 0.288 & 0.4 & 0.6 & 0.9921 & 0 & 0.8 \\
RIT:BBH:0251 & -7.50 & 5.00 & -4.76e-04 & 0.0811 & 0.3889 & 0.3666 & 0 & 0.288 & 0.4 & 0.6 & 0.9921 & 0 & 0.8 \\
RIT:BBH:0252 & -6.86 & 5.14 & -4.43e-04 & 0.08022 & 0.3704 & 0.2954 & 0.09184 & 0.2776 & 0.4286 & 0.5714 & 0.9912 & 0.5 & 0.85 \\
RIT:BBH:0253 & -4.00 & 4.00 & 0 & 0.104 & 0.5 & 0.5 & 0.2375 & 0.2375 & 0.5156 & 0.5156 & 0.9883 & 0.95 & 0.95 \\
RIT:BBH:0254 & -7.03 & 5.97 & -4.35e-04 & 0.08135 & 0.3718 & 0.5298 & 0.1267 & 0 & 0.4595 & 0.5405 & 0.9919 & 0.6 & 0 \\
RIT:BBH:0255 & -6.69 & 4.46 & -5.65e-04 & 0.08352 & 0.3878 & 0.3659 & 0 & 0.288 & 0.4 & 0.6 & 0.9908 & 0 & 0.8 \\
RIT:BBH:0256 & -6.69 & 4.46 & -5.66e-04 & 0.08354 & 0.3878 & 0.3659 & 0 & 0.288 & 0.4 & 0.6 & 0.9909 & 0 & 0.8 \\
RIT:BBH:0257 & -6.69 & 4.46 & -5.67e-04 & 0.08357 & 0.3878 & 0.3658 & 0 & 0.288 & 0.4 & 0.6 & 0.9909 & 0 & 0.8 \\
RIT:BBH:0258 & -6.69 & 4.46 & -5.68e-04 & 0.08359 & 0.3878 & 0.3658 & 0 & 0.288 & 0.4 & 0.6 & 0.9909 & 0 & 0.8 \\
RIT:BBH:0259 & -6.69 & 4.46 & -5.67e-04 & 0.08357 & 0.3878 & 0.3658 & 0 & 0.288 & 0.4 & 0.6 & 0.9909 & 0 & 0.8 \\
RIT:BBH:0260 & -6.69 & 4.46 & -5.66e-04 & 0.08354 & 0.3878 & 0.3659 & 0 & 0.288 & 0.4 & 0.6 & 0.9909 & 0 & 0.8 \\
RIT:BBH:0261 & -6.50 & 6.50 & -4.36e-04 & 0.08168 & 0.2579 & 0.4556 & 0.2125 & 0.1 & 0.5 & 0.5 & 0.992 & 0.85 & 0.4 \\
RIT:BBH:0262 & -6.86 & 5.14 & -4.64e-04 & 0.08124 & 0.3111 & 0.5392 & 0.1286 & 0.09796 & 0.4286 & 0.5714 & 0.9912 & 0.7 & 0.3 \\
RIT:BBH:0263 & -7.03 & 5.97 & -3.56e-04 & 0.07755 & 0.3985 & 0.2794 & 0.1056 & 0.2484 & 0.4595 & 0.5405 & 0.9917 & 0.5 & 0.85 \\
RIT:BBH:0264 & -7.05 & 4.70 & -5.25e-04 & 0.08267 & 0.3883 & 0.3662 & 0 & 0.288 & 0.4 & 0.6 & 0.9914 & 0 & 0.8 \\
RIT:BBH:0265 & -7.03 & 5.97 & -4.32e-04 & 0.08121 & 0.3983 & 0.2794 & 0.1056 & 0.2484 & 0.4595 & 0.5405 & 0.992 & 0.5 & 0.85 \\
RIT:BBH:0266 & -7.05 & 4.70 & -5.28e-04 & 0.08274 & 0.3883 & 0.3662 & 0 & 0.288 & 0.4 & 0.6 & 0.9915 & 0 & 0.8 \\
RIT:BBH:0267 & -7.05 & 4.70 & -5.33e-04 & 0.08287 & 0.3883 & 0.3661 & 0 & 0.288 & 0.4 & 0.6 & 0.9915 & 0 & 0.8 \\
RIT:BBH:0268 & -7.05 & 4.70 & -5.27e-04 & 0.08274 & 0.3883 & 0.3662 & 0 & 0.288 & 0.4 & 0.6 & 0.9915 & 0 & 0.8 \\
RIT:BBH:0269 & -6.84 & 4.56 & -5.45e-04 & 0.08314 & 0.388 & 0.366 & 0 & 0.288 & 0.4 & 0.6 & 0.9911 & 0 & 0.8 \\
RIT:BBH:0270 & -6.84 & 4.56 & -5.48e-04 & 0.08319 & 0.388 & 0.366 & 0 & 0.288 & 0.4 & 0.6 & 0.9911 & 0 & 0.8 \\
RIT:BBH:0271 & -6.84 & 4.56 & -5.52e-04 & 0.08329 & 0.388 & 0.3659 & 0 & 0.288 & 0.4 & 0.6 & 0.9912 & 0 & 0.8 \\
RIT:BBH:0272 & -6.84 & 4.56 & -5.54e-04 & 0.08334 & 0.388 & 0.3659 & 0 & 0.288 & 0.4 & 0.6 & 0.9912 & 0 & 0.8 \\
RIT:BBH:0273 & -6.84 & 4.56 & -5.52e-04 & 0.08329 & 0.388 & 0.3659 & 0 & 0.288 & 0.4 & 0.6 & 0.9912 & 0 & 0.8 \\
RIT:BBH:0274 & -6.84 & 4.56 & -5.47e-04 & 0.08319 & 0.388 & 0.366 & 0 & 0.288 & 0.4 & 0.6 & 0.9911 & 0 & 0.8 \\
RIT:BBH:0276 & -7.03 & 5.97 & -4.62e-04 & 0.08235 & 0.279 & 0.5162 & 0.1689 & 0.07305 & 0.4595 & 0.5405 & 0.9921 & 0.8 & 0.25 \\
RIT:BBH:0277 & -7.03 & 5.97 & -3.87e-04 & 0.07925 & 0.3985 & 0.2794 & 0.1056 & 0.2484 & 0.4595 & 0.5405 & 0.9918 & 0.5 & 0.85 \\
RIT:BBH:0278 & -10.83 & 2.17 & -1.15e-04 & 0.04369 & 0.09965 & 0.7352 & 0.02222 & 0.3472 & 0.1667 & 0.8333 & 0.9954 & 0.8 & 0.5 \\
RIT:BBH:0279 & -7.03 & 5.97 & -4.62e-04 & 0.08235 & 0.3983 & 0.4705 & 0.1056 & 0.1461 & 0.4595 & 0.5405 & 0.992 & 0.5 & 0.5 \\
RIT:BBH:0280 & -7.05 & 4.70 & -5.34e-04 & 0.08287 & 0.3883 & 0.3661 & 0 & 0.288 & 0.4 & 0.6 & 0.9915 & 0 & 0.8 \\
RIT:BBH:0281 & -7.05 & 4.70 & -5.36e-04 & 0.08293 & 0.3883 & 0.3661 & 0 & 0.288 & 0.4 & 0.6 & 0.9916 & 0 & 0.8 \\
RIT:BBH:0283 & -7.20 & 4.80 & -6.30e-04 & 0.08574 & 0.2415 & 0.3663 & 0.128 & 0.288 & 0.4 & 0.6 & 0.9922 & 0.8 & 0.8 \\
RIT:BBH:0284 & -9.00 & 3.00 & -3.91e-04 & 0.0671 & 0.2136 & 0.3911 & 0.03125 & 0.4781 & 0.25 & 0.75 & 0.9938 & 0.5 & 0.85 \\
RIT:BBH:0285 & -8.25 & 2.75 & -3.49e-04 & 0.06534 & 0.2132 & 0.3908 & 0.03125 & 0.4781 & 0.25 & 0.75 & 0.9928 & 0.5 & 0.85 \\
RIT:BBH:0286 & -9.00 & 3.00 & -3.64e-04 & 0.06616 & 0.2137 & 0.3912 & 0.03125 & 0.4781 & 0.25 & 0.75 & 0.9937 & 0.5 & 0.85 \\
RIT:BBH:0287 & -7.20 & 4.80 & -5.25e-04 & 0.0826 & 0.2415 & 0.3664 & 0.128 & 0.288 & 0.4 & 0.6 & 0.9918 & 0.8 & 0.8 \\
RIT:BBH:0288 & -8.25 & 2.75 & -3.35e-04 & 0.06449 & 0.2132 & 0.3908 & 0.03125 & 0.4781 & 0.25 & 0.75 & 0.9927 & 0.5 & 0.85 \\
RIT:BBH:0289 & -7.20 & 4.80 & -4.73e-04 & 0.08083 & 0.2416 & 0.3664 & 0.128 & 0.288 & 0.4 & 0.6 & 0.9916 & 0.8 & 0.8 \\
RIT:BBH:0290 & -5.71 & 4.29 & -1.40e-03 & 0.1082 & 0.4286 & 0.5714 & 0.1745 & 0.3102 & 0.4286 & 0.5714 & 0.9894 & 0.95 & 0.95 \\
RIT:BBH:0291 & -7.20 & 4.80 & -4.34e-04 & 0.07908 & 0.3451 & 0.4661 & 0.08 & 0.234 & 0.4 & 0.6 & 0.9913 & 0.5 & 0.65 \\
RIT:BBH:0292 & -7.20 & 4.80 & -4.18e-04 & 0.07823 & 0.2417 & 0.3664 & 0.128 & 0.288 & 0.4 & 0.6 & 0.9914 & 0.8 & 0.8 \\
RIT:BBH:0293 & -7.20 & 4.80 & -4.71e-04 & 0.08074 & 0.3451 & 0.466 & 0.08 & 0.234 & 0.4 & 0.6 & 0.9915 & 0.5 & 0.65 \\
RIT:BBH:0294 & -7.20 & 4.80 & -5.41e-04 & 0.08319 & 0.2416 & 0.5889 & 0.128 & 0 & 0.4 & 0.6 & 0.9918 & 0.8 & 0 \\
RIT:BBH:0295 & -7.20 & 4.80 & -4.33e-04 & 0.07899 & 0.2417 & 0.5232 & 0.128 & 0.18 & 0.4 & 0.6 & 0.9914 & 0.8 & 0.5 \\
RIT:BBH:0296 & -8.67 & 4.33 & -4.00e-04 & 0.07477 & 0.287 & 0.3466 & 0.05556 & 0.3778 & 0.3333 & 0.6667 & 0.9931 & 0.5 & 0.85 \\
RIT:BBH:0297 & -7.20 & 4.80 & -4.96e-04 & 0.08169 & 0.2416 & 0.5231 & 0.128 & 0.18 & 0.4 & 0.6 & 0.9917 & 0.8 & 0.5 \\
RIT:BBH:0298 & -7.20 & 4.80 & -4.58e-04 & 0.08014 & 0.3451 & 0.5739 & 0.08 & 0.09 & 0.4 & 0.6 & 0.9914 & 0.5 & 0.25 \\
RIT:BBH:0299 & -9.19 & 3.81 & -3.51e-04 & 0.0698 & 0.2517 & 0.3684 & 0.04289 & 0.425 & 0.2929 & 0.7071 & 0.9936 & 0.5 & 0.85 \\
RIT:BBH:0300 & -9.19 & 3.81 & -3.26e-04 & 0.06877 & 0.2517 & 0.3684 & 0.04289 & 0.425 & 0.2929 & 0.7071 & 0.9934 & 0.5 & 0.85 \\
RIT:BBH:0301 & -7.20 & 4.80 & -5.94e-04 & 0.08476 & 0.2415 & 0.523 & 0.128 & 0.18 & 0.4 & 0.6 & 0.992 & 0.8 & 0.5 \\
RIT:BBH:0302 & -7.20 & 4.80 & -5.47e-04 & 0.08339 & 0.345 & 0.5738 & 0.08 & 0.09 & 0.4 & 0.6 & 0.9917 & 0.5 & 0.25 \\
RIT:BBH:0303 & -7.29 & 5.21 & -4.14e-04 & 0.07919 & 0.2522 & 0.5726 & 0.1389 & 0 & 0.4167 & 0.5833 & 0.9917 & 0.8 & 0 \\
RIT:BBH:0304 & -7.20 & 4.80 & -5.28e-04 & 0.08271 & 0.3449 & 0.466 & 0.08 & 0.234 & 0.4 & 0.6 & 0.9917 & 0.5 & 0.65 \\
RIT:BBH:0305 & -7.20 & 4.80 & -5.02e-04 & 0.08188 & 0.345 & 0.5738 & 0.08 & 0.09 & 0.4 & 0.6 & 0.9916 & 0.5 & 0.25 \\
RIT:BBH:0306 & -7.20 & 4.80 & -5.90e-04 & 0.08465 & 0.3449 & 0.4659 & 0.08 & 0.234 & 0.4 & 0.6 & 0.9919 & 0.5 & 0.65 \\
RIT:BBH:0307 & -7.20 & 4.80 & -4.63e-04 & 0.08032 & 0.2416 & 0.589 & 0.128 & 0 & 0.4 & 0.6 & 0.9915 & 0.8 & 0 \\
RIT:BBH:0308 & -7.20 & 4.80 & -4.99e-04 & 0.08172 & 0.2416 & 0.5231 & 0.128 & 0.18 & 0.4 & 0.6 & 0.9917 & 0.8 & 0.5 \\
RIT:BBH:0309 & -7.20 & 4.80 & -4.94e-04 & 0.08154 & 0.345 & 0.5739 & 0.08 & 0.09 & 0.4 & 0.6 & 0.9915 & 0.5 & 0.25 \\
RIT:BBH:0311 & -7.20 & 4.80 & -4.61e-04 & 0.0803 & 0.3886 & 0.5231 & 0 & 0.18 & 0.4 & 0.6 & 0.9914 & 0 & 0.5 \\
RIT:BBH:0312 & -7.20 & 4.80 & -5.42e-04 & 0.0832 & 0.3884 & 0.5231 & 0 & 0.18 & 0.4 & 0.6 & 0.9917 & 0 & 0.5 \\
RIT:BBH:0314 & -8.67 & 4.33 & -2.82e-04 & 0.06922 & 0.2873 & 0.3467 & 0.05556 & 0.3778 & 0.3333 & 0.6667 & 0.9925 & 0.5 & 0.85 \\
RIT:BBH:0316 & -9.19 & 3.81 & -2.55e-04 & 0.06527 & 0.2519 & 0.3684 & 0.04289 & 0.425 & 0.2929 & 0.7071 & 0.9931 & 0.5 & 0.85 \\
RIT:BBH:0317 & -5.71 & 4.29 & -7.48e-04 & 0.09202 & 0.4286 & 0.5714 & 0.1745 & 0.3102 & 0.4286 & 0.5714 & 0.988 & 0.95 & 0.95 \\
RIT:BBH:0318 & -10.83 & 2.17 & -1.64e-04 & 0.04711 & 0.1422 & 0.4366 & 0.01389 & 0.5903 & 0.1667 & 0.8333 & 0.9957 & 0.5 & 0.85 \\
RIT:BBH:0319 & -7.29 & 5.21 & -4.83e-04 & 0.08199 & 0.2522 & 0.5725 & 0.1389 & 0 & 0.4167 & 0.5833 & 0.992 & 0.8 & 0 \\
RIT:BBH:0321 & -10.83 & 2.17 & -1.58e-04 & 0.04672 & 0.1422 & 0.4366 & 0.01389 & 0.5903 & 0.1667 & 0.8333 & 0.9957 & 0.5 & 0.85 \\
RIT:BBH:0322 & -9.19 & 3.81 & -2.45e-04 & 0.06445 & 0.2519 & 0.3684 & 0.04289 & 0.425 & 0.2929 & 0.7071 & 0.993 & 0.5 & 0.85 \\
RIT:BBH:0324 & -5.50 & 5.50 & -5.85e-04 & 0.08541 & 0.1802 & 0.1802 & 0.225 & 0.225 & 0.5 & 0.5 & 0.9904 & 0.9 & 0.9 \\
RIT:BBH:0336 & -8.67 & 4.33 & -3.48e-04 & 0.07276 & 0.2011 & 0.6573 & 0.08889 & 0 & 0.3333 & 0.6667 & 0.9929 & 0.8 & 0 \\
RIT:BBH:0337 & -10.83 & 2.17 & -1.11e-04 & 0.0434 & 0.1424 & 0.4366 & 0.01389 & 0.5903 & 0.1667 & 0.8333 & 0.9953 & 0.5 & 0.85 \\
RIT:BBH:0338 & -8.67 & 4.33 & -3.11e-04 & 0.07092 & 0.2011 & 0.6574 & 0.08889 & 0 & 0.3333 & 0.6667 & 0.9927 & 0.8 & 0 \\
RIT:BBH:0339 & -9.75 & 3.25 & -2.79e-04 & 0.06278 & 0.2413 & 0.4618 & 0 & 0.45 & 0.25 & 0.75 & 0.9941 & 0 & 0.8 \\
RIT:BBH:0344 & -7.81 & 4.69 & -5.07e-04 & 0.08096 & 0.3233 & 0.3241 & 0.07031 & 0.332 & 0.375 & 0.625 & 0.9925 & 0.5 & 0.85 \\
RIT:BBH:0345 & -10.83 & 2.17 & -1.09e-04 & 0.04311 & 0.1424 & 0.4366 & 0.01389 & 0.5903 & 0.1667 & 0.8333 & 0.9953 & 0.5 & 0.85 \\
RIT:BBH:0348 & -7.81 & 4.69 & -4.52e-04 & 0.0791 & 0.3641 & 0.5459 & 0 & 0.1953 & 0.375 & 0.625 & 0.9921 & 0 & 0.5 \\
RIT:BBH:0350 & -7.81 & 4.69 & -4.88e-04 & 0.08035 & 0.3233 & 0.4863 & 0.07031 & 0.2539 & 0.375 & 0.625 & 0.9923 & 0.5 & 0.65 \\
RIT:BBH:0352 & -7.81 & 4.69 & -4.82e-04 & 0.08013 & 0.364 & 0.3241 & 0 & 0.332 & 0.375 & 0.625 & 0.9923 & 0 & 0.85 \\
\end{longtable*}

\clearpage
\begin{longtable*}{lcccrrrrr}
\caption{The mass and spin of the nonprecessing BHBs in Table~\ref{tab:ID} after the
BHs had time to equilibrate ($t/m=200$).  Also provided are the difference in the masses,
$\delta m = (q-1)/(q+1)$, sum of the spins $S/m^2 = (\chi_{2z} + q^2\chi_{1z})/(q+1)^2$, and spin difference, 
$\Delta/m^2 = (\chi_{2z} - q\chi_{1z})/(1+q)$.}
\label{tab:IDr}\\
\hline
\hline
Run & $q^r$ & $m^r_1/m$ & $m^r_2/m$ & $\chi^r_{1z}$ & $\chi^r_{2z}$ & $\delta m_r$ & $S_r/m^2_r$ & $\Delta_r/m^2_r$ \\
\hline
\endfirsthead

\multicolumn{9}{c}
{{\tablename\ \thetable{} -- continued from previous page}}\\
\hline
Run & $q^r$ & $m^r_1/m$ & $m^r_2/m$ & $\chi^r_{1z}$ & $\chi^r_{2z}$ & $\delta m_r$ & $S_r/m^2_r$ & $\Delta_r/m^2_r$ \\
\hline
\endhead
\hline \multicolumn{9}{r}{{Continued on next page}} \\
\hline
\endfoot
\hline\hline
\endlastfoot

RIT:BBH:0136 &	$0.5001$ & $0.3333$ & $0.6666$ & $0.0000$ & $0.4244$ & $-0.3333$ & $0.2829$ & $0.1886$\\
RIT:BBH:0220 &	$0.6670$ & $0.4000$ & $0.5997$ & $0.0000$ & $-0.8008$ & $-0.1998$ & $-0.4801$ & $-0.2880$\\
RIT:BBH:0221 &	$0.2000$ & $0.1666$ & $0.8333$ & $-0.8005$ & $-0.5000$ & $-0.6667$ & $-0.2833$ & $-0.3694$\\
RIT:BBH:0222 &	$0.2000$ & $0.1666$ & $0.8333$ & $-0.8006$ & $0.5000$ & $-0.6667$ & $0.5500$ & $0.3250$\\
RIT:BBH:0223 &	$0.8496$ & $0.4593$ & $0.5405$ & $-0.8008$ & $0.2500$ & $-0.0813$ & $0.5028$ & $-0.0958$\\
RIT:BBH:0224 &	$0.2001$ & $0.1666$ & $0.8329$ & $0.8006$ & $0.8009$ & $-0.6666$ & $0.5334$ & $0.5778$\\
RIT:BBH:0226 &	$0.2000$ & $0.1666$ & $0.8333$ & $0.8006$ & $0.0000$ & $-0.6667$ & $-0.1334$ & $0.0222$\\
RIT:BBH:0227 &	$0.8496$ & $0.4593$ & $0.5405$ & $0.8007$ & $-0.2500$ & $-0.0813$ & $-0.5028$ & $0.0958$\\
RIT:BBH:0228 &	$0.9993$ & $0.4997$ & $0.5000$ & $0.8512$ & $-0.4000$ & $-0.0003$ & $-0.6251$ & $0.1125$\\
RIT:BBH:0232 &	$0.8500$ & $0.4593$ & $0.5403$ & $-0.8007$ & $-0.8008$ & $-0.0811$ & $-0.0649$ & $-0.4027$\\
RIT:BBH:0237 &	$0.7499$ & $0.4285$ & $0.5714$ & $-0.7003$ & $-0.3000$ & $-0.1429$ & $0.1286$ & $-0.2265$\\
RIT:BBH:0238 &	$0.7505$ & $0.4286$ & $0.5710$ & $-0.5000$ & $-0.8512$ & $-0.1425$ & $-0.2717$ & $-0.3694$\\
RIT:BBH:0239 &	$0.7505$ & $0.4286$ & $0.5710$ & $0.5000$ & $-0.8512$ & $-0.1425$ & $-0.7001$ & $-0.1857$\\
RIT:BBH:0240 &	$0.8506$ & $0.4595$ & $0.5402$ & $-0.5000$ & $-0.8508$ & $-0.0807$ & $-0.2297$ & $-0.3538$\\
RIT:BBH:0241 &	$0.7450$ & $0.4267$ & $0.5728$ & $-0.8972$ & $0.7723$ & $-0.1461$ & $0.8248$ & $0.0900$\\
RIT:BBH:0245 &	$0.5004$ & $0.3333$ & $0.6662$ & $0.5000$ & $-0.8514$ & $-0.3330$ & $-0.7335$ & $-0.3223$\\
RIT:BBH:0246 &	$0.2000$ & $0.1666$ & $0.8333$ & $-0.8004$ & $0.0000$ & $-0.6667$ & $0.1334$ & $-0.0222$\\
RIT:BBH:0247 &	$0.5004$ & $0.3333$ & $0.6662$ & $-0.5000$ & $0.8513$ & $-0.3330$ & $0.7334$ & $0.3222$\\
RIT:BBH:0249 &	$0.8499$ & $0.4594$ & $0.5405$ & $0.6001$ & $0.0000$ & $-0.0811$ & $-0.2757$ & $0.1267$\\
RIT:BBH:0252 &	$0.7505$ & $0.4286$ & $0.5710$ & $0.5000$ & $0.8512$ & $-0.1425$ & $0.2717$ & $0.3694$\\
RIT:BBH:0253 &	$1.0000$ & $0.5010$ & $0.5010$ & $0.9485$ & $0.9485$ & $0.0000$ & $0.0000$ & $0.4761$\\
RIT:BBH:0254 &	$0.8499$ & $0.4594$ & $0.5405$ & $-0.6001$ & $0.0000$ & $-0.0811$ & $0.2757$ & $-0.1267$\\
RIT:BBH:0261 &	$0.9993$ & $0.4997$ & $0.5000$ & $-0.8512$ & $0.4000$ & $-0.0003$ & $0.6251$ & $-0.1125$\\
RIT:BBH:0262 &	$0.7499$ & $0.4285$ & $0.5714$ & $0.7003$ & $0.3000$ & $-0.1429$ & $-0.1286$ & $0.2265$\\
RIT:BBH:0263 &	$0.8506$ & $0.4595$ & $0.5401$ & $0.5000$ & $0.8505$ & $-0.0807$ & $0.2296$ & $0.3537$\\
RIT:BBH:0265 &	$0.8506$ & $0.4595$ & $0.5402$ & $0.5000$ & $-0.8507$ & $-0.0807$ & $-0.6890$ & $-0.1427$\\
RIT:BBH:0276 &	$0.8496$ & $0.4593$ & $0.5405$ & $-0.8008$ & $-0.2500$ & $-0.0813$ & $0.2326$ & $-0.2419$\\
RIT:BBH:0277 &	$0.8506$ & $0.4595$ & $0.5401$ & $-0.5000$ & $0.8506$ & $-0.0807$ & $0.6889$ & $0.1426$\\
RIT:BBH:0278 &	$0.2000$ & $0.1666$ & $0.8333$ & $0.8006$ & $0.5001$ & $-0.6667$ & $0.2833$ & $0.3694$\\
RIT:BBH:0279 &	$0.8500$ & $0.4595$ & $0.5405$ & $-0.5001$ & $-0.5001$ & $-0.0811$ & $-0.0406$ & $-0.2517$\\
RIT:BBH:0283 &	$0.6667$ & $0.3999$ & $0.5997$ & $-0.8007$ & $-0.8006$ & $-0.2000$ & $-0.1599$ & $-0.4160$\\
RIT:BBH:0284 &	$0.3336$ & $0.2500$ & $0.7494$ & $-0.5001$ & $-0.8513$ & $-0.4997$ & $-0.5127$ & $-0.5094$\\
RIT:BBH:0285 &	$0.3336$ & $0.2500$ & $0.7494$ & $-0.5001$ & $0.8513$ & $-0.4997$ & $0.7626$ & $0.4469$\\
RIT:BBH:0286 &	$0.3336$ & $0.2500$ & $0.7494$ & $0.5000$ & $-0.8513$ & $-0.4997$ & $-0.7626$ & $-0.4469$\\
RIT:BBH:0287 &	$0.6667$ & $0.3998$ & $0.5997$ & $0.8007$ & $-0.8008$ & $-0.2000$ & $-0.8001$ & $-0.1600$\\
RIT:BBH:0288 &	$0.3336$ & $0.2500$ & $0.7494$ & $0.5001$ & $0.8513$ & $-0.4997$ & $0.5127$ & $0.5094$\\
RIT:BBH:0289 &	$0.6667$ & $0.3998$ & $0.5997$ & $-0.8007$ & $0.8008$ & $-0.2000$ & $0.8000$ & $0.1600$\\
RIT:BBH:0290 &	$0.7509$ & $0.4282$ & $0.5703$ & $-0.9510$ & $-0.9531$ & $-0.1422$ & $-0.1361$ & $-0.4844$\\
RIT:BBH:0291 &	$0.6668$ & $0.4000$ & $0.5999$ & $0.5001$ & $0.6503$ & $-0.1999$ & $0.1901$ & $0.3140$\\
RIT:BBH:0292 &	$0.6667$ & $0.3998$ & $0.5997$ & $0.8007$ & $0.8007$ & $-0.2000$ & $0.1600$ & $0.4160$\\
RIT:BBH:0293 &	$0.6668$ & $0.4000$ & $0.5999$ & $-0.5001$ & $0.6503$ & $-0.1999$ & $0.5901$ & $0.1540$\\
RIT:BBH:0294 &	$0.6664$ & $0.3998$ & $0.6000$ & $-0.8007$ & $0.0000$ & $-0.2002$ & $0.3201$ & $-0.1280$\\
RIT:BBH:0295 &	$0.6664$ & $0.3998$ & $0.6000$ & $0.8007$ & $0.5001$ & $-0.2002$ & $-0.0201$ & $0.3080$\\
RIT:BBH:0296 &	$0.5004$ & $0.3333$ & $0.6662$ & $-0.5000$ & $-0.8513$ & $-0.3330$ & $-0.4003$ & $-0.4334$\\
RIT:BBH:0297 &	$0.6664$ & $0.3998$ & $0.6000$ & $-0.8008$ & $0.5001$ & $-0.2002$ & $0.6201$ & $0.0520$\\
RIT:BBH:0298 &	$0.6667$ & $0.4000$ & $0.6000$ & $0.5001$ & $0.2500$ & $-0.2000$ & $-0.0500$ & $0.1700$\\
RIT:BBH:0299 &	$0.4145$ & $0.2929$ & $0.7066$ & $-0.5001$ & $-0.8513$ & $-0.4139$ & $-0.4548$ & $-0.4679$\\
RIT:BBH:0300 &	$0.4145$ & $0.2929$ & $0.7066$ & $0.5001$ & $-0.8513$ & $-0.4139$ & $-0.7476$ & $-0.3821$\\
RIT:BBH:0301 &	$0.6664$ & $0.3998$ & $0.6000$ & $-0.8007$ & $-0.5001$ & $-0.2002$ & $0.0201$ & $-0.3080$\\
RIT:BBH:0302 &	$0.6667$ & $0.4000$ & $0.6000$ & $-0.5001$ & $-0.2500$ & $-0.2000$ & $0.0500$ & $-0.1700$\\
RIT:BBH:0303 &	$0.7140$ & $0.4165$ & $0.5833$ & $0.8007$ & $0.0000$ & $-0.1669$ & $-0.3334$ & $0.1389$\\
RIT:BBH:0304 &	$0.6668$ & $0.4000$ & $0.5999$ & $0.5000$ & $-0.6503$ & $-0.1999$ & $-0.5901$ & $-0.1540$\\
RIT:BBH:0305 &	$0.6667$ & $0.4000$ & $0.6000$ & $-0.5001$ & $0.2500$ & $-0.2000$ & $0.3500$ & $0.0100$\\
RIT:BBH:0306 &	$0.6668$ & $0.4000$ & $0.5999$ & $-0.5001$ & $-0.6503$ & $-0.1999$ & $-0.1901$ & $-0.3140$\\
RIT:BBH:0307 &	$0.6664$ & $0.3998$ & $0.6000$ & $0.8007$ & $0.0000$ & $-0.2002$ & $-0.3201$ & $0.1280$\\
RIT:BBH:0308 &	$0.6664$ & $0.3998$ & $0.6000$ & $0.8007$ & $-0.5001$ & $-0.2002$ & $-0.6201$ & $-0.0520$\\
RIT:BBH:0309 &	$0.6667$ & $0.4000$ & $0.6000$ & $0.5000$ & $-0.2500$ & $-0.2000$ & $-0.3500$ & $-0.0100$\\
RIT:BBH:0311 &	$0.6667$ & $0.4000$ & $0.6000$ & $0.0000$ & $0.5001$ & $-0.2000$ & $0.3000$ & $0.1800$\\
RIT:BBH:0312 &	$0.6667$ & $0.4000$ & $0.6000$ & $-0.0000$ & $-0.5001$ & $-0.2000$ & $-0.3000$ & $-0.1800$\\
RIT:BBH:0314 &	$0.5004$ & $0.3333$ & $0.6662$ & $0.5000$ & $0.8513$ & $-0.3330$ & $0.4003$ & $0.4334$\\
RIT:BBH:0316 &	$0.4145$ & $0.2929$ & $0.7066$ & $-0.5001$ & $0.8513$ & $-0.4139$ & $0.7476$ & $0.3821$\\
RIT:BBH:0317 &	$0.7509$ & $0.4283$ & $0.5704$ & $0.9510$ & $0.9532$ & $-0.1422$ & $0.1361$ & $0.4845$\\
RIT:BBH:0318 &	$0.2002$ & $0.1667$ & $0.8327$ & $-0.5000$ & $-0.8514$ & $-0.6664$ & $-0.6252$ & $-0.6042$\\
RIT:BBH:0319 &	$0.7140$ & $0.4165$ & $0.5833$ & $-0.8007$ & $0.0000$ & $-0.1669$ & $0.3334$ & $-0.1389$\\
RIT:BBH:0321 &	$0.2002$ & $0.1667$ & $0.8327$ & $0.5000$ & $-0.8514$ & $-0.6664$ & $-0.7918$ & $-0.5764$\\
RIT:BBH:0322 &	$0.4145$ & $0.2929$ & $0.7066$ & $0.5001$ & $0.8513$ & $-0.4139$ & $0.4548$ & $0.4679$\\
RIT:BBH:0324 &	$1.0000$ & $0.4995$ & $0.4995$ & $0.9017$ & $0.9017$ & $0.0000$ & $-0.0000$ & $0.4500$\\
RIT:BBH:0336 &	$0.4998$ & $0.3332$ & $0.6667$ & $-0.8007$ & $0.0000$ & $-0.3335$ & $0.2668$ & $-0.0889$\\
RIT:BBH:0337 &	$0.2002$ & $0.1667$ & $0.8327$ & $-0.5001$ & $0.8514$ & $-0.6664$ & $0.7918$ & $0.5764$\\
RIT:BBH:0338 &	$0.4998$ & $0.3332$ & $0.6667$ & $0.8006$ & $0.0000$ & $-0.3335$ & $-0.2667$ & $0.0889$\\
RIT:BBH:0339 &	$0.3335$ & $0.2500$ & $0.7496$ & $-0.0000$ & $-0.8009$ & $-0.4998$ & $-0.6001$ & $-0.4500$\\
RIT:BBH:0344 &	$0.6004$ & $0.3750$ & $0.6246$ & $-0.5000$ & $-0.8513$ & $-0.2497$ & $-0.3440$ & $-0.4024$\\
RIT:BBH:0345 &	$0.2002$ & $0.1667$ & $0.8327$ & $0.5001$ & $0.8514$ & $-0.6664$ & $0.6252$ & $0.6042$\\
RIT:BBH:0348 &	$0.6000$ & $0.3750$ & $0.6250$ & $-0.0000$ & $-0.5001$ & $-0.2500$ & $-0.3125$ & $-0.1953$\\
RIT:BBH:0350 &	$0.6001$ & $0.3750$ & $0.6249$ & $-0.5000$ & $-0.6502$ & $-0.2499$ & $-0.2188$ & $-0.3242$\\
RIT:BBH:0352 &	$0.6004$ & $0.3750$ & $0.6245$ & $-0.0000$ & $-0.8513$ & $-0.2497$ & $-0.5314$ & $-0.3320$\\
\end{longtable*}

\clearpage
\begin{longtable*}{lcccrrrrrr}
\caption{The mass and spin of the precessing BHBs in Table~\ref{tab:ID} after the
BHs had time to equilibrate ($t/m=200$).}
\label{tab:IDr_prec}\\
\hline
\hline
Run & $q^r$ & $m^r_1/m$ & $m^r_2/m$ & $\chi^r_{1x}$ & $\chi^r_{1y}$ &  $\chi^r_{1z}$ &  $\chi^r_{2x}$ &  $\chi^r_{2y}$ & $\chi^r_{2z}$ \\
\hline
\endfirsthead

\multicolumn{10}{c}
{{\tablename\ \thetable{} -- continued from previous page}}\\
\hline
Run & $q^r$ & $m^r_1/m$ & $m^r_2/m$ & $\chi^r_{1x}$ & $\chi^r_{1y}$ &  $\chi^r_{1z}$ &  $\chi^r_{2x}$ &  $\chi^r_{2y}$ & $\chi^r_{2z}$ \\
\hline
\endhead
\hline \multicolumn{10}{r}{{Continued on next page}} \\
\hline
\endfoot
\hline\hline
\endlastfoot
RIT:BBH:0127 &	$0.7501$ & $0.4284$ & $0.5712$ & - & - & 0.8005 & 0.5115 & 0.1815 & -0.5887 \\
RIT:BBH:0128 &	$0.5001$ & $0.3332$ & $0.6663$ & 0.0124 & 0.0059 & 0.8004 & 0.6819 & 0.1550 & -0.3901 \\
RIT:BBH:0129 &	$0.3334$ & $0.2499$ & $0.7496$ & - & - & 0.8004 & 0.7473 & 0.1250 & -0.2596 \\
RIT:BBH:0130 &	$0.7500$ & $0.4284$ & $0.5712$ & - & - & -0.8007 & 0.5104 & 0.1674 & 0.5937 \\
RIT:BBH:0131 &	$0.5001$ & $0.3332$ & $0.6663$ & 0.0003 & -0.0055 & -0.8006 & 0.6686 & 0.1790 & 0.4025 \\
RIT:BBH:0132 &	$0.3334$ & $0.2499$ & $0.7496$ & 0.0097 & -0.0299 & -0.7999 & 0.7457 & 0.1224 & 0.2650 \\
RIT:BBH:0230 &	$0.6670$ & $0.4000$ & $0.5997$ & - & - & - & -0.5691 & -0.0579 & -0.5605 \\
RIT:BBH:0231 &	$0.6670$ & $0.4000$ & $0.5997$ & - & - & - & -0.4404 & -0.3553 & -0.5667 \\
RIT:BBH:0233 &	$0.6670$ & $0.4000$ & $0.5997$ & - & - & - & -0.2023 & -0.5351 & -0.5604 \\
RIT:BBH:0234 &	$0.6670$ & $0.4000$ & $0.5997$ & - & - & - & 0.0675 & -0.5711 & -0.5573 \\
RIT:BBH:0235 &	$0.6670$ & $0.4000$ & $0.5997$ & - & - & - & 0.3372 & -0.4621 & -0.5604 \\
RIT:BBH:0236 &	$0.6670$ & $0.4000$ & $0.5997$ & - & - & - & 0.5192 & -0.2284 & -0.5653 \\
RIT:BBH:0242 &	$0.6670$ & $0.4000$ & $0.5997$ & - & - & - & 0.4585 & 0.3360 & -0.5641 \\
RIT:BBH:0243 &	$0.6670$ & $0.4000$ & $0.5997$ & - & - & - & 0.1980 & 0.5253 & -0.5711 \\
RIT:BBH:0244 &	$0.6670$ & $0.4000$ & $0.5997$ & - & - & - & -0.5181 & 0.2330 & -0.5645 \\
RIT:BBH:0248 &	$0.6670$ & $0.4000$ & $0.5997$ & - & - & - & 0.5545 & 0.0630 & -0.5744 \\
RIT:BBH:0250 &	$0.6670$ & $0.4000$ & $0.5997$ & - & - & - & -0.0916 & 0.5607 & -0.5643 \\
RIT:BBH:0251 &	$0.6670$ & $0.4000$ & $0.5997$ & - & - & - & -0.3355 & 0.4590 & -0.5639 \\
RIT:BBH:0255 &	$0.6670$ & $0.4000$ & $0.5997$ & - & - & - & 0.3986 & 0.0379 & 0.6935 \\
RIT:BBH:0256 &	$0.6670$ & $0.4000$ & $0.5997$ & - & - & - & 0.3377 & 0.2299 & 0.6887 \\
RIT:BBH:0257 &	$0.6670$ & $0.4000$ & $0.5997$ & - & - & - & 0.1446 & 0.3677 & 0.6965 \\
RIT:BBH:0258 &	$0.6670$ & $0.4000$ & $0.5997$ & - & - & - & -0.0647 & 0.3934 & 0.6945 \\
RIT:BBH:0259 &	$0.6670$ & $0.4000$ & $0.5997$ & - & - & - & -0.2532 & 0.3151 & 0.6913 \\
RIT:BBH:0260 &	$0.6670$ & $0.4000$ & $0.5997$ & - & - & - & -0.3785 & 0.1469 & 0.6903 \\
RIT:BBH:0264 &	$0.6670$ & $0.4000$ & $0.5997$ & - & - & - & 0.7956 & 0.0910 & 0.0005 \\
RIT:BBH:0266 &	$0.6670$ & $0.4000$ & $0.5997$ & - & - & - & 0.6415 & 0.4794 & -0.0006 \\
RIT:BBH:0267 &	$0.6669$ & $0.4000$ & $0.5998$ & - & - & - & -0.5020 & 0.6238 & -0.0004 \\
RIT:BBH:0268 &	$0.6670$ & $0.4000$ & $0.5997$ & - & - & - & -0.7467 & 0.2894 & -0.0027 \\
RIT:BBH:0269 &	$0.6670$ & $0.4000$ & $0.5997$ & - & - & - & 0.6911 & 0.0897 & 0.3946 \\
RIT:BBH:0270 &	$0.6670$ & $0.4000$ & $0.5997$ & - & - & - & 0.5501 & 0.4193 & 0.4036 \\
RIT:BBH:0271 &	$0.6670$ & $0.4000$ & $0.5997$ & - & - & - & 0.2643 & 0.6419 & 0.3992 \\
RIT:BBH:0272 &	$0.6669$ & $0.4000$ & $0.5997$ & - & - & - & -0.1192 & 0.6815 & 0.4031 \\
RIT:BBH:0273 &	$0.6670$ & $0.4000$ & $0.5997$ & - & - & - & -0.4388 & 0.5414 & 0.3943 \\
RIT:BBH:0274 &	$0.6670$ & $0.4000$ & $0.5997$ & - & - & - & -0.6429 & 0.2683 & 0.3949 \\
RIT:BBH:0280 &	$0.6670$ & $0.4000$ & $0.5997$ & - & - & - & 0.2865 & 0.7477 & 0.0043 \\
RIT:BBH:0281 &	$0.6669$ & $0.4000$ & $0.5998$ & - & - & - & -0.1417 & 0.7880 & 0.0055 \\
\end{longtable*}

\clearpage
\begin{longtable*}{lcccc}
\caption{Table of the initial orbital frequency $m\omega_i$,
number of orbits to merger, $N$, and the initial and final eccentricities,
$e_i$ and $e_f$ for the spinning cases.}
\label{tab:ecc}\\
\hline
\hline
Run & $m\omega_i$ & $N$ & $e_i$ & $e_f$ \\
\hline
\endfirsthead

\multicolumn{5}{c}
{{\tablename\ \thetable{} -- continued from previous page}}\\
\hline
Run & $m\omega_i$ & $N$ & $e_i$ & $e_f$ \\
\hline
\endhead
\hline \multicolumn{5}{r}{{Continued on next page}} \\
\hline
\endfoot
\hline\hline
\endlastfoot

RIT:BBH:0127 &	$0.0317$ & $5.9$ & $0.0117$ & $0.0022$\\
RIT:BBH:0128 &	$0.0309$ & $6.3$ & $0.0108$ & $0.0033$\\
RIT:BBH:0129 &	$0.0300$ & $7.3$ & $0.0092$ & $0.0010$\\
RIT:BBH:0130 &	$0.0315$ & $6.3$ & $0.0117$ & $0.0013$\\
RIT:BBH:0131 &	$0.0306$ & $7.0$ & $0.0110$ & $0.0039$\\
RIT:BBH:0132 &	$0.0296$ & $8.2$ & $0.0097$ & $0.0044$\\
RIT:BBH:0136 &	$0.0156$ & $22.0$ & $0.0242$ & $0.0009$\\
RIT:BBH:0220 &	$0.0211$ & $8.8$ & $0.0064$ & $0.0024$\\
RIT:BBH:0221 &	$0.0199$ & $14.1$ & $0.0036$ & $0.0015$\\
RIT:BBH:0222 &	$0.0192$ & $23.6$ & $0.0026$ & $0.0006$\\
RIT:BBH:0223 &	$0.0196$ & $11.2$ & $0.0042$ & $0.0011$\\
RIT:BBH:0224 &	$0.0189$ & $29.6$ & $0.0024$ & $0.0006$\\
RIT:BBH:0226 &	$0.0194$ & $20.5$ & $0.0024$ & $0.0009$\\
RIT:BBH:0227 &	$0.0194$ & $13.6$ & $0.0041$ & $0.0011$\\
RIT:BBH:0228 &	$0.0194$ & $13.6$ & $0.0041$ & $0.0012$\\
RIT:BBH:0230 &	$0.0210$ & $9.6$ & $0.0063$ & $0.0012$\\
RIT:BBH:0231 &	$0.0210$ & $9.6$ & $0.0064$ & $0.0010$\\
RIT:BBH:0232 &	$0.0201$ & $8.3$ & $0.0045$ & $0.0016$\\
RIT:BBH:0233 &	$0.0209$ & $9.7$ & $0.0043$ & $0.0015$\\
RIT:BBH:0234 &	$0.0209$ & $9.7$ & $0.0043$ & $0.0011$\\
RIT:BBH:0235 &	$0.0209$ & $9.6$ & $0.0041$ & $0.0017$\\
RIT:BBH:0236 &	$0.0210$ & $9.6$ & $0.0039$ & $0.0014$\\
RIT:BBH:0237 &	$0.0226$ & $7.9$ & $0.0042$ & $0.0019$\\
RIT:BBH:0238 &	$0.0229$ & $6.8$ & $0.0039$ & $0.0020$\\
RIT:BBH:0239 &	$0.0224$ & $8.6$ & $0.0040$ & $0.0017$\\
RIT:BBH:0240 &	$0.0199$ & $8.9$ & $0.0042$ & $0.0014$\\
RIT:BBH:0241 &	$0.0306$ & $6.4$ & $0.0062$ & $0.0029$\\
RIT:BBH:0242 &	$0.0210$ & $9.6$ & $0.0063$ & $0.0012$\\
RIT:BBH:0243 &	$0.0209$ & $9.7$ & $0.0043$ & $0.0015$\\
RIT:BBH:0244 &	$0.0210$ & $9.6$ & $0.0039$ & $0.0019$\\
RIT:BBH:0245 &	$0.0198$ & $10.7$ & $0.0036$ & $0.0001$\\
RIT:BBH:0246 &	$0.0195$ & $18.5$ & $0.0032$ & $0.0009$\\
RIT:BBH:0247 &	$0.0192$ & $16.7$ & $0.0039$ & $0.0008$\\
RIT:BBH:0248 &	$0.0210$ & $9.6$ & $0.0038$ & $0.0012$\\
RIT:BBH:0249 &	$0.0193$ & $13.9$ & $0.0043$ & $0.0010$\\
RIT:BBH:0250 &	$0.0209$ & $9.7$ & $0.0043$ & $0.0017$\\
RIT:BBH:0251 &	$0.0209$ & $9.6$ & $0.0041$ & $0.0017$\\
RIT:BBH:0252 &	$0.0216$ & $14.0$ & $0.0041$ & $0.0013$\\
RIT:BBH:0253 &	$0.0419$ & $6.7$ & $0.0248$ & $0.0038$\\
RIT:BBH:0254 &	$0.0196$ & $10.9$ & $0.0045$ & $0.0012$\\
RIT:BBH:0255 &	$0.0245$ & $10.8$ & $0.0026$ & $0.0023$\\
RIT:BBH:0256 &	$0.0244$ & $10.8$ & $0.0027$ & $0.0016$\\
RIT:BBH:0257 &	$0.0244$ & $10.8$ & $0.0024$ & $0.0016$\\
RIT:BBH:0258 &	$0.0244$ & $10.8$ & $0.0024$ & $0.0016$\\
RIT:BBH:0259 &	$0.0244$ & $10.8$ & $0.0024$ & $0.0014$\\
RIT:BBH:0260 &	$0.0244$ & $10.8$ & $0.0025$ & $0.0014$\\
RIT:BBH:0261 &	$0.0196$ & $11.1$ & $0.0042$ & $0.0011$\\
RIT:BBH:0262 &	$0.0218$ & $12.6$ & $0.0028$ & $0.0009$\\
RIT:BBH:0263 &	$0.0191$ & $16.6$ & $0.0045$ & $0.0010$\\
RIT:BBH:0264 &	$0.0230$ & $9.9$ & $0.0038$ & $0.0013$\\
RIT:BBH:0265 &	$0.0196$ & $11.2$ & $0.0068$ & $0.0010$\\
RIT:BBH:0266 &	$0.0229$ & $9.9$ & $0.0036$ & $0.0019$\\
RIT:BBH:0267 &	$0.0228$ & $9.9$ & $0.0041$ & $0.0017$\\
RIT:BBH:0268 &	$0.0229$ & $9.9$ & $0.0032$ & $0.0013$\\
RIT:BBH:0269 &	$0.0239$ & $10.4$ & $0.0021$ & $0.0016$\\
RIT:BBH:0270 &	$0.0238$ & $10.4$ & $0.0039$ & $0.0019$\\
RIT:BBH:0271 &	$0.0237$ & $10.4$ & $0.0045$ & $0.0020$\\
RIT:BBH:0272 &	$0.0236$ & $10.4$ & $0.0036$ & $0.0018$\\
RIT:BBH:0273 &	$0.0237$ & $10.4$ & $0.0041$ & $0.0016$\\
RIT:BBH:0274 &	$0.0238$ & $10.4$ & $0.0036$ & $0.0013$\\
RIT:BBH:0276 &	$0.0198$ & $9.7$ & $0.0044$ & $0.0018$\\
RIT:BBH:0277 &	$0.0193$ & $14.0$ & $0.0067$ & $0.0011$\\
RIT:BBH:0278 &	$0.0191$ & $25.8$ & $0.0019$ & $0.0008$\\
RIT:BBH:0279 &	$0.0198$ & $9.7$ & $0.0045$ & $0.0019$\\
RIT:BBH:0280 &	$0.0227$ & $10.0$ & $0.0044$ & $0.0018$\\
RIT:BBH:0281 &	$0.0227$ & $10.0$ & $0.0044$ & $0.0024$\\
RIT:BBH:0283 &	$0.0231$ & $6.5$ & $0.0042$ & $0.0033$\\
RIT:BBH:0284 &	$0.0229$ & $7.5$ & $0.0029$ & $0.0020$\\
RIT:BBH:0285 &	$0.0245$ & $14.0$ & $0.0029$ & $0.0011$\\
RIT:BBH:0286 &	$0.0226$ & $8.5$ & $0.0026$ & $0.0012$\\
RIT:BBH:0287 &	$0.0224$ & $9.1$ & $0.0034$ & $0.0018$\\
RIT:BBH:0288 &	$0.0243$ & $15.3$ & $0.0035$ & $0.0010$\\
RIT:BBH:0289 &	$0.0220$ & $11.4$ & $0.0036$ & $0.0009$\\
RIT:BBH:0290 &	$0.0314$ & $3.4$ & $0.0192$ & $0.0083$\\
RIT:BBH:0291 &	$0.0217$ & $13.6$ & $0.0040$ & $0.0010$\\
RIT:BBH:0292 &	$0.0216$ & $14.8$ & $0.0040$ & $0.0012$\\
RIT:BBH:0293 &	$0.0219$ & $11.6$ & $0.0041$ & $0.0011$\\
RIT:BBH:0294 &	$0.0224$ & $8.8$ & $0.0042$ & $0.0015$\\
RIT:BBH:0295 &	$0.0217$ & $13.6$ & $0.0024$ & $0.0011$\\
RIT:BBH:0296 &	$0.0200$ & $9.1$ & $0.0037$ & $0.0014$\\
RIT:BBH:0297 &	$0.0221$ & $10.4$ & $0.0059$ & $0.0011$\\
RIT:BBH:0298 &	$0.0218$ & $12.1$ & $0.0042$ & $0.0011$\\
RIT:BBH:0299 &	$0.0200$ & $9.3$ & $0.0056$ & $0.0012$\\
RIT:BBH:0300 &	$0.0198$ & $10.7$ & $0.0057$ & $0.0010$\\
RIT:BBH:0301 &	$0.0228$ & $7.3$ & $0.0043$ & $0.0026$\\
RIT:BBH:0302 &	$0.0224$ & $8.6$ & $0.0069$ & $0.0018$\\
RIT:BBH:0303 &	$0.0205$ & $13.1$ & $0.0039$ & $0.0013$\\
RIT:BBH:0304 &	$0.0223$ & $9.1$ & $0.0067$ & $0.0014$\\
RIT:BBH:0305 &	$0.0221$ & $10.2$ & $0.0066$ & $0.0015$\\
RIT:BBH:0306 &	$0.0227$ & $7.4$ & $0.0066$ & $0.0023$\\
RIT:BBH:0307 &	$0.0219$ & $11.8$ & $0.0068$ & $0.0014$\\
RIT:BBH:0308 &	$0.0222$ & $10.1$ & $0.0069$ & $0.0016$\\
RIT:BBH:0309 &	$0.0221$ & $10.4$ & $0.0068$ & $0.0014$\\
RIT:BBH:0311 &	$0.0218$ & $12.0$ & $0.0042$ & $0.0011$\\
RIT:BBH:0312 &	$0.0224$ & $8.7$ & $0.0069$ & $0.0018$\\
RIT:BBH:0314 &	$0.0190$ & $18.6$ & $0.0040$ & $0.0008$\\
RIT:BBH:0316 &	$0.0191$ & $18.2$ & $0.0036$ & $0.0007$\\
RIT:BBH:0317 &	$0.0281$ & $10.8$ & $0.0066$ & $0.0010$\\
RIT:BBH:0318 &	$0.0201$ & $11.7$ & $0.0030$ & $0.0007$\\
RIT:BBH:0319 &	$0.0210$ & $9.7$ & $0.0042$ & $0.0014$\\
RIT:BBH:0321 &	$0.0200$ & $12.7$ & $0.0029$ & $0.0011$\\
RIT:BBH:0322 &	$0.0190$ & $20.0$ & $0.0039$ & $0.0008$\\
RIT:BBH:0324 &	$0.0247$ & $12.4$ & $0.0037$ & $0.0012$\\
RIT:BBH:0336 &	$0.0196$ & $12.0$ & $0.0060$ & $0.0009$\\
RIT:BBH:0337 &	$0.0189$ & $28.3$ & $0.0026$ & $0.0004$\\
RIT:BBH:0338 &	$0.0194$ & $14.9$ & $0.0035$ & $0.0012$\\
RIT:BBH:0339 &	$0.0199$ & $10.7$ & $0.0052$ & $0.0008$\\
RIT:BBH:0344 &	$0.0213$ & $7.8$ & $0.0062$ & $0.0011$\\
RIT:BBH:0345 &	$0.0188$ & $29.7$ & $0.0030$ & $0.0005$\\
RIT:BBH:0348 &	$0.0209$ & $9.8$ & $0.0066$ & $0.0014$\\
RIT:BBH:0350 &	$0.0212$ & $8.5$ & $0.0065$ & $0.0014$\\
RIT:BBH:0352 &	$0.0212$ & $8.7$ & $0.0062$ & $0.0016$\\
\end{longtable*}

\clearpage
\begin{longtable*}{lcr}
\caption{The energy radiated, $\delta \mathcal{M}^{IH} = M_{adm} - M_{rem}^{IH}$,
and final spin, $\chi_{\mathrm{rem}}^{IH}$, as measured using the IH formalism
The error bars are due to variations in the measured mass and spin with time.  
For aligned systems with final spin antialigned to the initial orbital angular 
momentum the minus sign is preserved.}
\label{tab:spinerad}\\
\hline
\hline
Run & $\delta \mathcal{M}^{IH}$ & $\chi_{\mathrm{rem}}^{IH}$ \\
\hline
\hline
\endfirsthead

\multicolumn{3}{c}
{{\tablename\ \thetable{} -- continued from previous page}}\\
\hline
Run & $\delta \mathcal{M}^{IH}$ & $\chi_{\mathrm{rem}}^{IH}$ \\
\hline
\endhead
\hline \multicolumn{3}{r}{{Continued on next page}} \\
\hline
\endfoot
\hline\hline
\endlastfoot
RIT:BBH:0127 &  $0.047049 \pm 0.000002$ & 	$0.642923 \pm 0.000004$ \\
RIT:BBH:0128 &  $0.039357 \pm 0.000006$ & 	$0.609307 \pm 0.000086$ \\
RIT:BBH:0129 &  $0.029840 \pm 0.000001$ & 	$0.607187 \pm 0.000012$ \\
RIT:BBH:0130 &  $0.051886 \pm 0.000001$ & 	$0.730998 \pm 0.000003$ \\
RIT:BBH:0131 &  $0.043854 \pm 0.000071$ & 	$0.768088 \pm 0.000570$ \\
RIT:BBH:0132 &  $0.034965 \pm 0.000000$ & 	$0.756259 \pm 0.000002$ \\
RIT:BBH:0136 &  $0.049301 \pm 0.000002$ & 	$0.775972 \pm 0.000008$ \\
RIT:BBH:0220 &  $0.034360 \pm 0.000000$ & 	$0.463421 \pm 0.000032$ \\
RIT:BBH:0221 &  $0.013239 \pm 0.000006$ & 	$0.104715 \pm 0.000006$ \\
RIT:BBH:0222 &  $0.024213 \pm 0.000002$ & 	$0.697263 \pm 0.000030$ \\
RIT:BBH:0223 &  $0.043177 \pm 0.000003$ & 	$0.627577 \pm 0.000000$ \\
RIT:BBH:0224 &  $0.039046 \pm 0.000140$ & 	$0.880312 \pm 0.005699$ \\
RIT:BBH:0226 &  $0.018361 \pm 0.000004$ & 	$0.427060 \pm 0.000001$ \\
RIT:BBH:0227 &  $0.053982 \pm 0.000003$ & 	$0.733438 \pm 0.000081$ \\
RIT:BBH:0228 &  $0.055817 \pm 0.000003$ & 	$0.751804 \pm 0.000196$ \\
RIT:BBH:0230 &  $0.038221 \pm 0.000002$ & 	$0.555080 \pm 0.000038$ \\
RIT:BBH:0231 &  $0.037993 \pm 0.000002$ & 	$0.556462 \pm 0.000038$ \\
RIT:BBH:0232 &  $0.033037 \pm 0.000001$ & 	$0.419364 \pm 0.000090$ \\
RIT:BBH:0233 &  $0.037722 \pm 0.000002$ & 	$0.558094 \pm 0.000038$ \\
RIT:BBH:0234 &  $0.037552 \pm 0.000002$ & 	$0.558906 \pm 0.000037$ \\
RIT:BBH:0235 &  $0.037696 \pm 0.000002$ & 	$0.557694 \pm 0.000038$ \\
RIT:BBH:0236 &  $0.038129 \pm 0.000001$ & 	$0.555337 \pm 0.000040$ \\
RIT:BBH:0237 &  $0.037119 \pm 0.000001$ & 	$0.531769 \pm 0.000006$ \\
RIT:BBH:0238 &  $0.033155 \pm 0.000002$ & 	$0.428593 \pm 0.000007$ \\
RIT:BBH:0239 &  $0.039182 \pm 0.000001$ & 	$0.542678 \pm 0.000004$ \\
RIT:BBH:0240 &  $0.034421 \pm 0.000003$ & 	$0.451168 \pm 0.000033$ \\
RIT:BBH:0241 &  $0.051719 \pm 0.000007$ & 	$0.743552 \pm 0.000049$ \\
RIT:BBH:0242 &  $0.037993 \pm 0.000002$ & 	$0.556462 \pm 0.000038$ \\
RIT:BBH:0243 &  $0.037722 \pm 0.000002$ & 	$0.558094 \pm 0.000038$ \\
RIT:BBH:0244 &  $0.038129 \pm 0.000001$ & 	$0.555336 \pm 0.000040$ \\
RIT:BBH:0245 &  $0.029791 \pm 0.000003$ & 	$0.374712 \pm 0.000000$ \\
RIT:BBH:0246 &  $0.016920 \pm 0.000004$ & 	$0.405685 \pm 0.000002$ \\
RIT:BBH:0247 &  $0.059225 \pm 0.000000$ & 	$0.854389 \pm 0.000011$ \\
RIT:BBH:0248 &  $0.038220 \pm 0.000002$ & 	$0.555080 \pm 0.000037$ \\
RIT:BBH:0249 &  $0.055860 \pm 0.000004$ & 	$0.755050 \pm 0.000049$ \\
RIT:BBH:0250 &  $0.037552 \pm 0.000002$ & 	$0.558906 \pm 0.000037$ \\
RIT:BBH:0251 &  $0.037696 \pm 0.000002$ & 	$0.557694 \pm 0.000037$ \\
RIT:BBH:0252 &  $0.079713 \pm 0.000012$ & 	$0.890918 \pm 0.000245$ \\
RIT:BBH:0253 &  $0.105833 \pm 0.000053$ & 	$0.940241 \pm 0.000027$ \\
RIT:BBH:0254 &  $0.041832 \pm 0.000003$ & 	$0.605990 \pm 0.000026$ \\
RIT:BBH:0255 &  $0.063541 \pm 0.000001$ & 	$0.832657 \pm 0.000032$ \\
RIT:BBH:0256 &  $0.063236 \pm 0.000002$ & 	$0.833626 \pm 0.000050$ \\
RIT:BBH:0257 &  $0.062808 \pm 0.000003$ & 	$0.834735 \pm 0.000094$ \\
RIT:BBH:0258 &  $0.062559 \pm 0.000003$ & 	$0.835152 \pm 0.000087$ \\
RIT:BBH:0259 &  $0.062811 \pm 0.000001$ & 	$0.834240 \pm 0.000049$ \\
RIT:BBH:0260 &  $0.063373 \pm 0.000003$ & 	$0.832833 \pm 0.000088$ \\
RIT:BBH:0261 &  $0.043225 \pm 0.000002$ & 	$0.615218 \pm 0.000023$ \\
RIT:BBH:0262 &  $0.062388 \pm 0.000001$ & 	$0.806245 \pm 0.000047$ \\
RIT:BBH:0263 &  $0.079755 \pm 0.000086$ & 	$0.885051 \pm 0.001694$ \\
RIT:BBH:0264 &  $0.048571 \pm 0.000002$ & 	$0.710881 \pm 0.000014$ \\
RIT:BBH:0265 &  $0.041759 \pm 0.000006$ & 	$0.584610 \pm 0.000056$ \\
RIT:BBH:0266 &  $0.047876 \pm 0.000000$ & 	$0.713698 \pm 0.000001$ \\
RIT:BBH:0267 &  $0.047019 \pm 0.000001$ & 	$0.715990 \pm 0.000006$ \\
RIT:BBH:0268 &  $0.048562 \pm 0.000002$ & 	$0.710338 \pm 0.000012$ \\
RIT:BBH:0269 &  $0.054878 \pm 0.000001$ & 	$0.792807 \pm 0.000012$ \\
RIT:BBH:0270 &  $0.055376 \pm 0.000001$ & 	$0.791199 \pm 0.000013$ \\
RIT:BBH:0271 &  $0.055820 \pm 0.000001$ & 	$0.790455 \pm 0.000008$ \\
RIT:BBH:0272 &  $0.056897 \pm 0.000000$ & 	$0.787993 \pm 0.000005$ \\
RIT:BBH:0273 &  $0.057247 \pm 0.000001$ & 	$0.787618 \pm 0.000011$ \\
RIT:BBH:0274 &  $0.055833 \pm 0.000000$ & 	$0.791237 \pm 0.000007$ \\
RIT:BBH:0276 &  $0.037611 \pm 0.000002$ & 	$0.530494 \pm 0.000037$ \\
RIT:BBH:0277 &  $0.057208 \pm 0.000003$ & 	$0.773654 \pm 0.000085$ \\
RIT:BBH:0278 &  $0.026856 \pm 0.000002$ & 	$0.715530 \pm 0.000043$ \\
RIT:BBH:0279 &  $0.037334 \pm 0.000002$ & 	$0.521499 \pm 0.000062$ \\
RIT:BBH:0280 &  $0.047518 \pm 0.000001$ & 	$0.715617 \pm 0.000012$ \\
RIT:BBH:0281 &  $0.046779 \pm 0.000002$ & 	$0.717849 \pm 0.000014$ \\
RIT:BBH:0283 &  $0.030805 \pm 0.000002$ & 	$0.383336 \pm 0.000002$ \\
RIT:BBH:0284 &  $0.019466 \pm 0.000003$ & 	$0.136838 \pm 0.000011$ \\
RIT:BBH:0285 &  $0.052236 \pm 0.000043$ & 	$0.883435 \pm 0.001081$ \\
RIT:BBH:0286 &  $0.020768 \pm 0.000001$ & 	$0.172724 \pm 0.000017$ \\
RIT:BBH:0287 &  $0.039013 \pm 0.000001$ & 	$0.539101 \pm 0.000036$ \\
RIT:BBH:0288 &  $0.060428 \pm 0.000149$ & 	$0.907461 \pm 0.001717$ \\
RIT:BBH:0289 &  $0.054364 \pm 0.000001$ & 	$0.781275 \pm 0.000012$ \\
RIT:BBH:0290 &  $0.030423 \pm 0.000016$ & 	$0.350609 \pm 0.000009$ \\
RIT:BBH:0291 &  $0.068688 \pm 0.000005$ & 	$0.854343 \pm 0.000115$ \\
RIT:BBH:0292 &  $0.083983 \pm 0.000212$ & 	$0.906306 \pm 0.001585$ \\
RIT:BBH:0293 &  $0.053790 \pm 0.000001$ & 	$0.773129 \pm 0.000003$ \\
RIT:BBH:0294 &  $0.039046 \pm 0.000002$ & 	$0.589673 \pm 0.000061$ \\
RIT:BBH:0295 &  $0.068155 \pm 0.000005$ & 	$0.844768 \pm 0.000009$ \\
RIT:BBH:0296 &  $0.026874 \pm 0.000004$ & 	$0.308819 \pm 0.000001$ \\
RIT:BBH:0297 &  $0.047266 \pm 0.000001$ & 	$0.711723 \pm 0.000013$ \\
RIT:BBH:0298 &  $0.055193 \pm 0.000001$ & 	$0.766060 \pm 0.000001$ \\
RIT:BBH:0299 &  $0.023398 \pm 0.000003$ & 	$0.233298 \pm 0.000026$ \\
RIT:BBH:0300 &  $0.025421 \pm 0.000009$ & 	$0.283357 \pm 0.000034$ \\
RIT:BBH:0301 &  $0.033411 \pm 0.000001$ & 	$0.462101 \pm 0.000007$ \\
RIT:BBH:0302 &  $0.037663 \pm 0.000001$ & 	$0.555664 \pm 0.000026$ \\
RIT:BBH:0303 &  $0.054631 \pm 0.000001$ & 	$0.746572 \pm 0.000005$ \\
RIT:BBH:0304 &  $0.038894 \pm 0.000000$ & 	$0.549270 \pm 0.000030$ \\
RIT:BBH:0305 &  $0.045062 \pm 0.000000$ & 	$0.679210 \pm 0.000010$ \\
RIT:BBH:0306 &  $0.033390 \pm 0.000001$ & 	$0.453122 \pm 0.000005$ \\
RIT:BBH:0307 &  $0.052619 \pm 0.000000$ & 	$0.732976 \pm 0.000018$ \\
RIT:BBH:0308 &  $0.043136 \pm 0.000001$ & 	$0.613587 \pm 0.000039$ \\
RIT:BBH:0309 &  $0.044675 \pm 0.000001$ & 	$0.648076 \pm 0.000015$ \\
RIT:BBH:0311 &  $0.055694 \pm 0.000003$ & 	$0.781539 \pm 0.000019$ \\
RIT:BBH:0312 &  $0.037596 \pm 0.000001$ & 	$0.540226 \pm 0.000029$ \\
RIT:BBH:0314 &  $0.073484 \pm 0.000049$ & 	$0.903065 \pm 0.000334$ \\
RIT:BBH:0316 &  $0.056812 \pm 0.000009$ & 	$0.871082 \pm 0.000187$ \\
RIT:BBH:0317 &  $0.104701 \pm 0.000080$ & 	$0.942466 \pm 0.000058$ \\
RIT:BBH:0318 &  $0.011757 \pm 0.000002$ & 	$-0.105807 \pm 0.000008$ \\
RIT:BBH:0319 &  $0.039589 \pm 0.000000$ & 	$0.589102 \pm 0.000053$ \\
RIT:BBH:0321 &  $0.012213 \pm 0.000001$ & 	$-0.090347 \pm 0.000006$ \\
RIT:BBH:0322 &  $0.068114 \pm 0.000014$ & 	$0.906331 \pm 0.000200$ \\
RIT:BBH:0324 &  $0.099686 \pm 0.000022$ & 	$0.927790 \pm 0.001526$ \\
RIT:BBH:0336 &  $0.035055 \pm 0.000000$ & 	$0.574463 \pm 0.000013$ \\
RIT:BBH:0337 &  $0.037908 \pm 0.000002$ & 	$0.895267 \pm 0.000044$ \\
RIT:BBH:0338 &  $0.043459 \pm 0.000001$ & 	$0.669149 \pm 0.000004$ \\
RIT:BBH:0339 &  $0.020426 \pm 0.000001$ & 	$0.178243 \pm 0.000006$ \\
RIT:BBH:0344 &  $0.030016 \pm 0.000001$ & 	$0.370895 \pm 0.000005$ \\
RIT:BBH:0345 &  $0.041385 \pm 0.000010$ & 	$0.904325 \pm 0.000217$ \\
RIT:BBH:0348 &  $0.035551 \pm 0.000001$ & 	$0.513676 \pm 0.000012$ \\
RIT:BBH:0350 &  $0.031769 \pm 0.000000$ & 	$0.428983 \pm 0.000007$ \\
RIT:BBH:0352 &  $0.031913 \pm 0.000001$ & 	$0.414030 \pm 0.000008$ \\
\end{longtable*}


\bibliographystyle{apsrev4-1}
\bibliography{../../Bibtex/references}

\end{document}